\documentclass{aa}  

\usepackage{graphicx}
\usepackage[dvipsnames]{xcolor}
\usepackage[normalem]{ulem}
\usepackage{array}
\usepackage{txfonts}
\newcommand{\teff}{$T_{\mathrm{eff}}$}
\newcommand{\msun}{M$_\odot$}
\newcommand{\logg}{$\log g$}
\newcommand\omicron{o}
\newcommand{\AJ}[1]{\textcolor{black}{#1}}%{\textcolor{green}{#1}} 
\newcommand{\shreeya}[1]{\textcolor{black}{#1}}%{\textcolor{blue}{#1}} 
\renewcommand\ion[2]{{#1\;}{\scshape{#2}}}%                       % ion, i.e., CII = \ion{C}{ii}
\renewcommand{\deg}{$^\circ$}
\renewcommand\omicron{o}
\newcommand{\omiori}{$\omicron^1$~Ori}

\begin{document}

   \title{S stars and s-process in the \textit{Gaia} era}
      \subtitle{II. Constraining the luminosity of the third dredge-up with Tc-rich S stars}
% with instead of via?

   \author{Shreeya Shetye
          \inst{1,2}
          \and
          Sophie Van Eck\inst{1} \and
          Alain Jorissen \inst{1} \and
          Stephane Goriely\inst{1}    \and      
          Lionel Siess\inst{1}   \and
          Hans Van Winckel\inst{2} \and
          Bertrand Plez \inst{3}\and
          Michel Godefroid \inst{4} \and
          George Wallerstein \inst{5}
          }

   \institute{ Institute of Astronomy and Astrophysics (IAA), Université libre de Bruxelles (ULB)
,
              CP  226,  Boulevard  du  Triomphe,  B-1050  Bruxelles, Belgium\\
              \email{Shreeya.Shetye@ulb.ac.be}
         \and
             Institute of Astronomy, KU Leuven, Celestijnenlaan 200D, B-3001 Leuven, Belgium
        \and
        Laboratoire Univers et Particules de Montpellier, Université de Montpellier, CNRS 34095, Montpellier Cedex 05, France
        \and
        Spectroscopy, Quantum Chemistry and Atmospheric Remote Sensing (SQUARES), CP160/09, Universit\'e libre de Bruxelles (ULB), 1050 Brussels, Belgium
        \and
        Department of Astronomy, University of Washington, Box 351580, Seattle, WA 98195-1580, USA 
             }

   \date{Received; accepted }

% \abstract{}{}{}{}{} 
% 5 {} token are mandatory
 
  \abstract
  % context heading (optional)
   {S stars are late-type giants that are transition objects between M-type stars and carbon stars on the asymptotic giant branch (AGB). They are classified into two types: intrinsic or extrinsic, based on the presence or absence of technetium (Tc). 
   %\LSrm{}{Tc is an s-process element with no stable isotope.}
   The Tc-rich or intrinsic S stars are thermally-pulsing (TP-)AGB stars internally producing  s-process elements (including Tc) which are brought to their surface via the third dredge-up (TDU). Tc-poor or extrinsic S stars gained 
   %owe 
   their s-process overabundances 
   %to a \SVE{mass transfer from a } binary 
   %interaction 
   via accretion of s-process-rich material from an AGB companion which has since turned into a dim white dwarf. } 
  % aims heading (mandatory)
   {Our goal is to investigate the evolutionary status of Tc-rich S stars by locating them in a Hertzsprung-Russell (HR) diagram using the results of Gaia
   early Data Release 3 (EDR3).
   %Gaia Data Release 2 (DR2). 
    We combine 
   %the abundances derived from 
   the current sample of 13 Tc-rich stars with %those from 
   our previous studies of 10 Tc-rich stars to determine the observational onset of the TDU 
   in the metallicity range [-0.7; 0]. 
   We also %confront 
   compare our 
   %derived 
   abundance determinations with dedicated AGB nucleosynthesis predictions.
   %of AGB models. 
  }
  % methods heading (mandatory)
   {The stellar parameters are derived using an iterative tool which combines HERMES high-resolution spectra, accurate Gaia EDR3 parallaxes, stellar evolution models and tailored MARCS model atmospheres for S-type stars. Using these stellar parameters we determine the heavy-element abundances by line synthesis. } 
   {In the HR diagram, the intrinsic S stars are located at higher luminosities than 
   the predicted onset of the TDU.
   These findings are consistent with 
   Tc-rich S stars being genuinely TP-AGB stars. 
   The comparison of the derived s-process abundance profiles of our intrinsic S stars with the nucleosynthesis predictions provide an overall good agreement. 
   Stars with highest [s/Fe] tend to have the highest C/O ratios. 
   }
   {}


   \maketitle
%
%________________________________________________________________

\section{Introduction}
\label{sec:intro}

S stars are late-type giants displaying, as the most characteristic feature of their optical spectra,
ZrO bands  \citep{Merrill} along with the usual TiO bands present in stars of similar temperatures (approximately 3000-4000~K, as M-type stars). The spectra show overabundances of s-process elements \citep{smithlambertfeb1990} produced by the slow capture of neutrons on elements heavier than Fe \citep{burbidge1957, s-process} 
during the thermally-pulsing AGB phase (TP-AGB).
These elements are then brought to the surface of the AGB star through mixing processes called third dredge-ups (TDU). The carbon over oxygen (C/O) ratio of S stars is in the range 0.5 to 0.99  which makes them transition objects between M-type (C/O~$\sim$~0.4) and carbon (C/O >1) stars \citep{IbenRenzini}.

Another important characteristic of the S star family is the technetium (Tc) dichotomy. Tc is an element produced by the s-process and which has no stable isotope. Its isotope $^{99}$Tc has a half-life of 210~000 yrs. The puzzling detection of Tc in some S stars but not in others was solved after realizing that the S stars without Tc ({\it i.e.} Tc-poor S stars) 
belong to binary systems. They are called {\it extrinsic} S stars \citep{smith1988, jorissen1993} 
because they owe their s-process abundances to mass transfer %scenario with 
from a former AGB companion which is now a white dwarf. Therefore, the extrinsic S stars show 
s-process overabundances, except for Tc, which has decayed since the 
termination of the AGB phase of the companion. On the contrary, Tc-rich S stars, called {\it intrinsic}, are producing s-process elements including Tc and transporting them to their surface via ongoing, recurrent  TDU episodes. 

We recently discovered a third class of S stars, the {\it bitrinsic} S stars \cite[S20 hereafter]{bitrinsic} that share properties with both the intrinsic and extrinsic classes:
they are Tc-rich and as such are located on the TP-AGB, but they also show overabundances of Nb, a signature of extrinsically-enriched stars \citep{pieter2015, shreeya1, Drisya}. Indeed, the unstable $^{93}$Zr isotope (produced by the s-process) decays into $^{93}$Nb, the only stable isotope of Nb, in 1.53~Myr. Therefore a niobium enrichment is expected in extrinsic stars, while it is not in TP-AGB stars.

The intrinsic S stars play an important role in our understanding of AGB nucleosynthesis and in particular of the TDU. They are the first objects on the AGB to show clear signatures (Tc and ZrO) of the TDU 
occurrences. They constrain the minimum luminosity of the TDU as well as its mass and metallicity dependence. The recent discovery of intrinsic, low-mass (initial mass < 1.5 \msun), solar metallicity
S stars  \cite[S19 hereafter]{shreeya2} provided an observational evidence of the operation of TDUs even in low-mass stars. 

In a previous paper \citep[S18 hereafter]{shreeya1} we introduced a new method 
for the parameter and abundance determination in S-type stars.
 However, the number of intrinsic S stars available at the time of that study was limited 
 by the bias of the Tycho-Gaia astrometric solution against extremely red sources \citep{michalik}.
In the current work we investigate the chemical composition and evolutionary status of 13 intrinsic S stars from Gaia DR2 for which we could obtain high-resolution spectra. This paper is a follow up on the investigation of the evolutionary status of S stars conducted in the pioneering work of \cite{smithlambertfeb1990}.
We introduce the observational sample in Section~\ref{observationalsample} 
and discuss the Tc detection in Section~\ref{Tclines}. Sections~\ref{parametersdetermination} and \ref{adundance determination} are dedicated to the description of the parameter and abundance determination.
In Section~\ref{sectHRD} we present the Gaia EDR3 %\citep{gdr2} 
HR diagram of intrinsic and extrinsic S stars.
We discuss the derived elemental abundances of the sample stars and compare 
them with nucleosynthesis predictions in Section~\ref{sectabundances}. 
Finally, we conclude with a summary of the most important results. 

%__________________________________________________________________

\section{Observational sample}\label{observationalsample}

From the General Catalogue of Galactic S-Stars \citep[CGSS]{cgss}, we selected the ones with a 
Gaia Data Release 2 (DR2) parallax matching the condition $\sigma_{\varpi} / \varpi$~$\leq 0.3$
and with available HERMES high-resolution spectra \citep{raskin}. From this sample we distinguished the intrinsic S stars from the extrinsic ones using Tc lines (Section~\ref{Tclines}) and kept only 
the Tc-rich S stars. Among these, the Tc-rich S stars with initial masses smaller than 1.5~\msun~were studied in S19. The cases of BD~+79\deg156 and $\omicron^1$ Ori, two bitrinsic stars,
were discussed in S20. In the current work we derive parameters and abundances for
the remaining Tc-rich S stars and also compile the Tc-rich S star results from S18 (3 stars), S19 (5 stars), S20 (2 stars) to increase the size of our sample. 

Though the sample was designed using the Gaia DR2, Gaia early Data Release 3 (EDR3) became public during the course of our analysis. Hence the stellar parameters and luminosities were derived using Gaia EDR3. The relative parallax difference between the two releases $|\varpi_{\rm DR2} - \varpi_{\rm EDR3}|/ \varpi_{\rm EDR3}$ is usually smaller than 12\% (except for V812~Oph, CSS~151, and CSS~454 which show the largest relative deviations of 22\%, 33\% and 57.5\%, respectively).  The cause of the large parallax difference of some stars between DR2 and EDR3 is not yet clear. However, to make a consistent  comparison of the present study with the previously studied samples (S18, S19 and S20), we re-computed the luminosities of all the extrinsic and intrinsic stars of S18, S19 and S20 using EDR3 (Table~\ref{parametersrevised}). As a result, some stars previously identified (using DR2 parallaxes) as low-mass S stars (BD +34$^\circ$~1698 = CSS~413 and HD~357941 = CSS~1190; see S19) turn out to have higher masses (M$_{\rm ini}$ = 1.4 and 3.5~M$_\odot$ respectively) according to their location in the HR diagram using EDR3 parallaxes (Sect.~\ref{sectHRD}). This change in their masses required to adopt a \logg~value different from the one used for their abundance analysis in S19 (both stars had \logg\ $\sim$ 1 in S19; with EDR3 we find instead \logg\ $\sim$ 0 for both of them). Hence, in the current work we  also re-computed the stellar parameters and abundances of BD +34$^\circ$1698 and HD~357941, along the guidelines described in Sections.~\ref{parametersdetermination} and \ref{adundance determination} using the EDR3 parallaxes (Table~\ref{abundtablelowmass}).

A last criterion to define the Tc-rich S-star sample relates to the photometric variability. 
Large convective cells and/or pulsations are responsible for the photometric variability of TP-AGB stars. 
The thermal structure of these pulsating atmospheres 
can be significantly different from 
time-independent, hydrostatic atmospheres used to derive their stellar parameters, which can therefore be quite unreliable.
We checked the time series data obtained by the American Association of Variable Star Observers (AAVSO) and removed from our sample the Tc-rich S stars with photometric variability $\Delta V >$ 1 mag (see Table~\ref{photometricvar}) except for V812 Oph. The HERMES spectra of V812 Oph, despite its high variability ($\Delta V >$ 1.4 mag), matched well the atmospheric models all over the optical range, hence we included it in the current sample.
The  sample of newly analyzed stars is listed in Table~\ref{basic data}.

\begin{table}[]
    \centering
    \caption{The Tc-rich S stars 
 excluded from the current study because of their large photometric variability. 
 }
    \label{photometricvar}
    \begin{tabular}{rcc}
    \hline
          CGSS & AAVSO & $\Delta V$ (mag) \\
          \hline
          9 & 0018 + 38 & 7.65 \\
          307 & 0701 + 22 & 7.53 \\
          347 & 0720 + 46 & 1.39 \\       
          856 & 1425 + 84 & 2.95 \\
          1219 & 2023 + 36 & 1.10 \\
          1115 & 1910 - 07 & 4.76 \\
          1309 & 2245 +17 & 4.74 \\
          \hline
    \end{tabular}
     \tablefoot{Columns 1 and 2 list the CGSS \citep[]{cgss} 
 and AAVSO identifiers, respectively. Column 3 contains the $V$ amplitude from AAVSO.}
\end{table}

\begin{table*}
\caption{\label{basic data} Basic data for our S star sample. }
\centering
\renewcommand{\arraystretch}{1.5}
\begin{tabular}{r l l c c c c c c c c c c}
\hline
%\rule{0pt}{20pt}
 CSS & Name & Sp. type & $V$ & $K$ & $J$ & \multicolumn{1}{c}{$\varpi_{\rm EDR3}\pm\sigma_{\varpi}$} & $Z_5$ & $E_{B-V}$ & $BC_{K}$ & Observation date & S/N \\
  & & & & & & (mas) & (mas) & & & & \\
\hline
89           & V1139 Tau&S & 7.90  & 0.97  & 2.27  & 1.79 $\pm$ 0.05 & -0.047 & 0.34 & 3.00   & 23 November 2012 & 150       \\
151          & CSS 151    &S  & 10.99 & 4.64  & 5.97  &0.36 $\pm$ 0.03 & -0.053 & 0.27 & 2.91   &31 July 2017 &    40   \\
233          & HD 288833 &S3/2  & 9.40  & 4.26  & 5.51  & 0.83 $\pm$ 0.02 & -0.038 & 0.31 & 2.84  &20 August 2012 &      100   \\
265          & KR CMa & M4S    & 8.50  & 1.96  & 3.30  & 1.29 $\pm$ 0.03 & -0.030 & 0.11& 3.01    &10 April 2011 & 70        \\
312          & AA Cam & M5S    & 7.79  & 1.39  & 2.59  & 2.08 $\pm$ 0.05 & -0.024& 0.13 & 2.84   & 9 April 2011 & 90        \\
416          & HD 64147&S5/2.5 & 9.20   & 3.61  & 5.00  & 0.56 $\pm$ 0.03 & -0.041 & 0.09 & 2.93 & 27 November 2016 &40\\
454          & CSS 454 &S  & 10.39  & 4.93  & 6.13  & 0.40 $\pm$ 0.03 & -0.045 & 0.07 & 2.91 & 2 February 2017 & 50 \\
474          & CD -27\deg5131 &S4,2 & 9.62  & 2.72  & 4.35  & 0.64 $\pm$ 0.03 &-0.031 & 0.14& 3.01 & 2 April 2010 & 70    \\
597          & BD -18\deg2608 &S & 9.58   & 4.01  & 5.27  & 0.71 $\pm$ 0.02 &-0.038& 0.14 & 2.91 & 24 March 2018 & 70   \\
997          & V812 Oph&S5+/2.5  & 10.58 & 4.00  & 5.22  & 0.64 $\pm$ 0.03 & -0.053 & 0.22 & 2.93  & 22 April 2016 & 40         \\
1070 &  V679 Oph & S5/5+ & 8.92         & 2.69 & 4.16  & 1.04 $\pm$ 0.03 & -0.049 & 0.37 & 2.83 & 16 March 2018 & 80        \\
1152 & Vy 12 & S5-/6        & 10.13        & 3.76 & 5.06  & 0.58 $\pm$ 0.02 & -0.037 & 0.21 & 2.99 &  18 March 2018 & 80   \\
1315 	& HR Peg &S4+/1+&	 6.36 &	 0.94 &	 2.15 &	 2.14 $\pm$	 0.08 & -0.043& 0.12 & 2.93 & 17 July 2009 &   60      \\
\hline
\end{tabular}
\tablefoot{Columns 1 and 2 list different identifiers: CGSS \citep[]{cgss} entry number and HD, BD or other names. Columns 3, 4, 5, and 6 indicate spectral type, $V$, $K$ and $J$ magnitudes, respectively, extracted from the SIMBAD Astronomical Database \citep{simbad}. Column 7 lists the Gaia EDR3 parallax and its error. $Z_5$ is the five-parameter zero-point correction on the parallax from \cite{lindegrenzpt2020}. The reddenning $E_{B-V}$ has been obtained from the \cite{gontcharov2017} extinction maps. $BC_K$ is the bolometric correction in the $K$ band as computed from the MARCS model atmospheres. Columns 10 and 11 document the HERMES  observation dates 
%of the HERMES spectra 
and their signal-to-noise (S/N) ratio around 500 nm.}
\end{table*}
\renewcommand{\arraystretch}{1}

\section{Tc detection}\label{Tclines}

\begin{figure*}[!htbp]     
\begin{centering}
    \mbox{\includegraphics[scale=0.31]{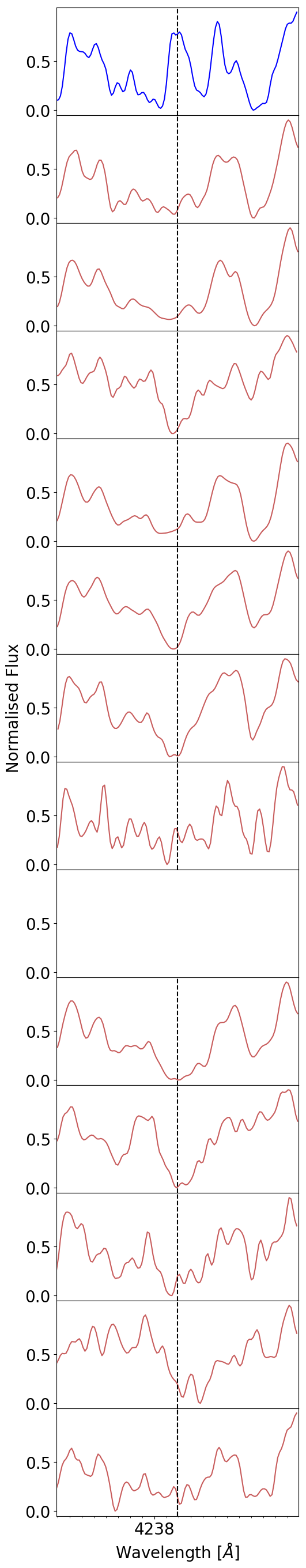}}   
    \hspace{0.1px}
    \mbox{\includegraphics[scale=0.31]{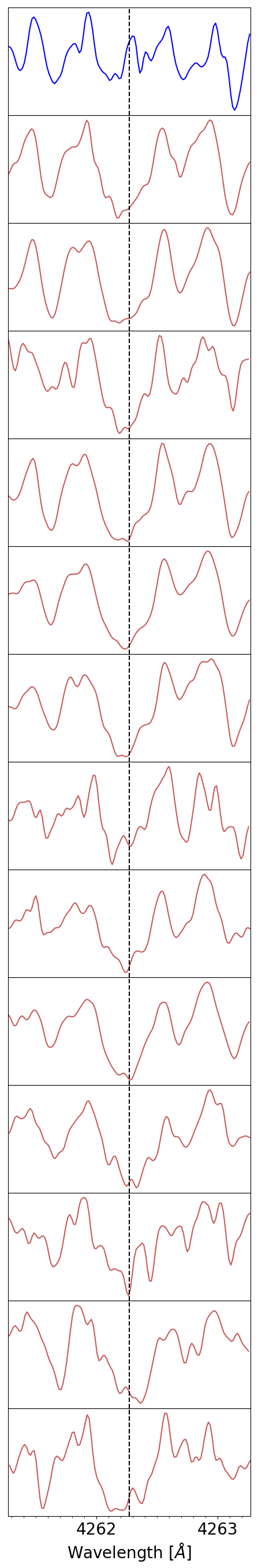}}
    \hspace{0.1px}
    \mbox{\includegraphics[scale=0.31]{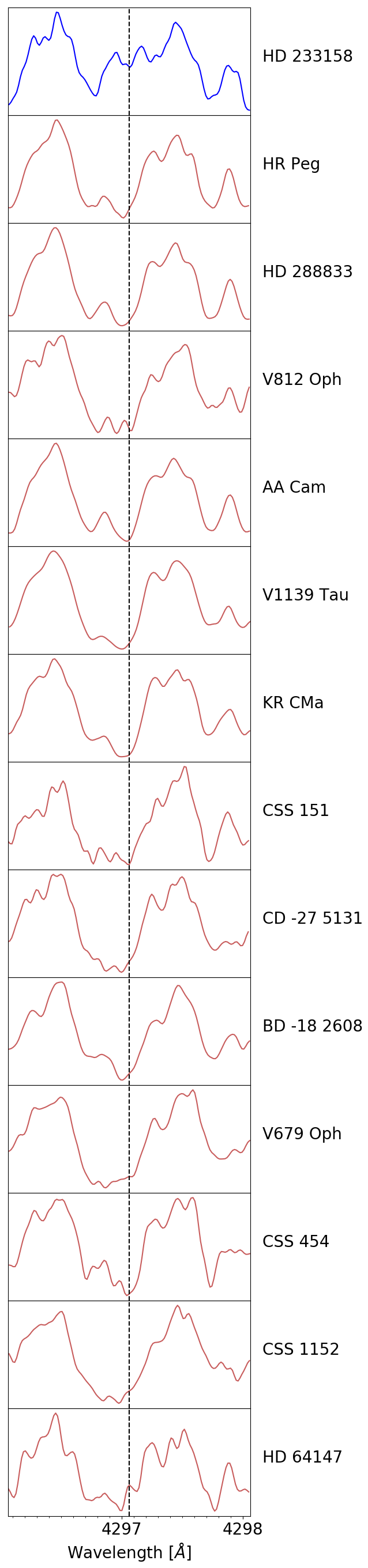}}
    \caption{\label{Tc1}
    The spectral region around the three
    (4238.19, 4262.27 and 4297.06~\AA) \ion{Tc}{I} lines in our sample of Tc-rich S stars. For comparison purposes, the spectrum of a Tc-poor S star (HD 233158, in blue in the top  panels) from S18 is also plotted. The spectra have been arbitrarily normalized and binned by a factor of 1.5 to increase the S/N ratio. The \ion{Tc}{I} line at 4238.19~\AA~could not be used for CD~-27\deg5131 as local normalization was hampered by a nearby cosmic ray hit.}
\end{centering}
\end{figure*}

Three \ion{Tc}{I} resonance lines located at 4238.19, 4262.27 and 4297.06~\AA~were used. 
A signal to noise ratio (S/N) of at least 30 in the $V$-band is needed to detect the Tc lines in S-star spectra \citep[S18, S19]{hipp}. The spectra of our sample stars all have a S/N 
(listed in Column 11 of Table~\ref{basic data})
equal to or larger than 40 in the $V$ band and  30 in the $B$ band. The Tc absorption features of our sample stars around the three Tc lines are presented in Fig.~\ref{Tc1}.

We confirm all previous Tc-rich classifications except for one target, HD~288833, classified as Tc-poor by \cite{jorissen1993}, despite the fact that its IRAS excess ([12] - [25]) is larger than $-1.3$
while extrinsic stars have [12] - [25]~$< -1.3$ according to their classification criterion. This criterion 
was designed by \cite{jorissen1993} to diagnose the expected lower infrared excess of the extrinsic stars with respect to that of (more evolved) intrinsic stars. The three Tc absorption features can be readily identified for this object in  Fig.~\ref{Tc1}.

For all other stars our classification is in agreement with 
previous findings when they exist. HR Peg, AA Cam, V1139 Tau, V679 Oph and HD~64147 were indeed classified as intrinsic S stars in \cite{smith1988}, \cite{smithlambertfeb1990} and \cite{jorissen1993}.
We agree with \cite{otto} classification of CSS~1152 as an intrinsic S star based on AKARI photometry
and also confirm the results of \cite{pieter2015} that KR CMa and CSS 454 are intrinsic S stars. 
We could not use the Tc line at 4238.19~\AA~for CD -27$^\circ$5131, as the region around this line is dominated by a cosmic ray hit. From its other two Tc lines we assess the Tc-rich nature of this object, in agreement with SVE17. For V812 Oph, CSS~151 and BD-18$^\circ$2608, that we classify as Tc-rich, no literature classification is available.

%===============================================

\section{Derivation of the atmospheric parameters}\label{parametersdetermination}

Disentangling the intricate parameter space of S stars has always been a challenging task because their spectra are dominated by molecules. The atmospheres of S stars are complex as the thermal structure is dependent on their chemical composition (C/O and, to a lesser extent, heavy elements). Different methods have been used in the past for the stellar parameters determination. For instance, in their pioneering work,
\cite{smithandlambert1985} derived \teff~from the $(V-K)$ colors and \logg~from the standard \logg~- mass relationship where they estimated the masses by comparing the positions of the S stars in the HR diagram with the evolutionary tracks. However, the \teff~- $(V-K)$ relation valid for M-type giants has been shown to be inappropriate for S-type stars (SVE17) because of the spectral energy distribution (SED) alteration (mainly by ZrO, LaO and YO bands) induced by the non-standard chemical composition of S stars.
New MARCS model atmospheres \citep{gustafsson2008} were designed to cover the full parameter range of S stars, with \teff~from 2700~K to 4000~K, \logg~from 0 to 5, [Fe/H] of 0.0 and -0.5, C/O 
values of 0.500, 0.752, 0.899, 0.925, 0.951, 0.971, 0.991 and [s/Fe] of 0.0, 1.0 and 2.0 (SVE17).

In the current study, the stellar parameters are determined as in S18. In summary, the high-resolution HERMES spectra are compared with a grid of $\sim 3500$ synthetic spectra computed from the S star MARCS models to obtain atmospheric parameter estimates \teff, log~$g$ and [Fe/H]. 
The fitting is undertaken over small spectral windows \shreeya{(listed in Appendix Table~\ref{Tab:bands})} 
because HERMES spectra cannot be considered as \AJ{flux-calibrated}  over their whole wavelength range. 
The fit with the lowest total $\chi^2$ value then identifies the best-fitting model, providing a first estimate of the atmospheric parameters. 

The stellar luminosities were calculated using the distances derived from the Gaia EDR3 parallaxes 
after applying the zero-point correction from \cite{lindegrenzpt2020}, the reddening $E_{B-V}$ computed from \cite{gontcharov2017} and the bolometric correction in the $K$ band as computed from the MARCS model atmospheres. Combining the \teff, metallicity and luminosity, we located the stars in the HR diagram and compared them with STAREVOL \citep{siess2006} evolutionary tracks to estimate their current stellar masses. 
 
A new surface gravity was then computed and the procedure iterated (as described in Fig.~5 of S18)
until the \logg~ retrieved from spectral fitting was consistent with the one obtained from the 
HR diagram. The uncertainties on the stellar parameters were obtained from the variations of the atmospheric parameters while iterating for \logg. 
Our final set of parameters and the corresponding uncertainties are presented in Table~\ref{finalparams}. 
A further discussion of the reliability of the stellar masses thus found, based on their correlation with the height above the galactic plane and with 2MASS photometry, may be found in Appendix~\ref{massqualitycheck}.

\begin{table*}
\caption{\label{finalparams} Atmospheric parameters for S stars.}
\centering %
\setlength\tabcolsep{4pt}
\begin{tabular}{l c c  c l l c c c c}
\hline 
 Name & $T_{\rm eff}$ & $L$ &  $\log g$ & [Fe/H] &$\sigma_{\rm [Fe/H]}$& C/O & [s/Fe] & M$_{\rm curr}$ & M$_{\rm ini}$\\
  & (K) & ($L_{\odot}$) & (dex) & & & &(dex) &($M_{\odot}$)&($M_{\odot}$) \\
\hline \\
V1139 Tau & 3400 & 6600 & 1 & -0.06 (13) & 0.13 & 0.752 & 1 &  3.3 & 3.5\\
 & (3400 - 3500)  & (6200 - 6900) & (1 - 3)  & & & (0.500 - 0.752)  &(1 - 1)  & &  \\
CSS 151  & 3500 & 4600 & 1 & -0.25 (14) & 0.14 & 0.500  & 1  & 2.6 & 3.2\\
 & (3500 - 3600)  & (4100 - 5200) & (1 - 3)  & & & (0.500 - 0.899) &(1 - 1)  & &  \\
HD 288833 & 3600  & 1700 & 1 & -0.30 (13) & 0.14 & 0.500 & 0  &  1.3 & 1.4 \\
 & (3600 - 3600)  & (1600 - 1800) & (1 - 2) & & & (0.500 - 0.752)  & (0 - 1) & &  \\
 KR CMa & 3400 & 4800 & 1  & -0.01 (13) & 0.22 & 0.500 & 1  & 2.8 & 3.0\\
 & (3400 - 3500)  & (4600 - 5000) & (1 - 3)  & & & (0.500 - 0.752) &(1 - 1)  & &  \\
AA Cam   & 3600 & 3700  & 1 & -0.32 (15) & 0.13 & 0.500  & 0  & 2.4 & 2.5 \\
 & (3600 - 3600) & (3600 - 3900) & (1 - 2)  & & & (0.500 - 0.899)  &(0 - 1)  & &  \\
HD 64147 & 3500  & 5400  & 0 & -0.67 (11) & 0.10 & 0.500  & 0 & 2.7 & 2.9\\
 &(3500 - 3600)  & (4900 - 5900) &(0 - 1)  & & & (0.500 - 0.752) &(0 - 1)  & &  \\
CSS 454 & 3500   & 2900 & 1 & -0.38 (12) & 0.14 & 0.500  & 1 & 1.8 & 1.9 \\
 & (3500 - 3600) & (2600 - 3400) & (1 - 3)  & & & (0.500 - 0.899) & (1 - 1)  & &  \\
CD -27\deg5131 & 3400 & 9200 & 0 & -0.30 (9) & 0.12 & 0.500 & 1 & 3.4 & 3.7\\
  &(3400 - 3600))  & (8300 - 10100) & (0 - 3) & & & (0.500 - 0.899)  &(1 - 1)  & &  \\
BD -18\deg2608 & 3500 & 2400 & 1 & -0.31 (13) & 0.16 & 0.752  & 1 & 1.5 & 1.6\\
 &(3500 - 3600)  & (2300 - 2600) & (1 - 3)  & & & (0.500 - 0.899)  &(1 - 1)  & &  \\
V812 Oph & 3500 & 3000 & 1  & -0.37 (13) & 0.13 & 0.500 & 1 & 1.8& 2.0\\
  & (3500 - 3500)  & (2700 - 3200) & (1 - 2)  & & & (0.500 - 0.899) &(1 - 1)  & &  \\
V679 Oph   & 3600  & 4600 & 1 & -0.52 (9)  & 0.14 &  0.899 & 1 & 2.0 & 2.1\\
 &(3600 - 3600)  & (4300 - 4800) & (1 - 1) & & & (0.752 - 0.899)  &(1 - 2)  & &  \\ 
CSS 1152 & 3400  & 4400 & 1 & -0.14 (6) &0.13 &0.971  & 1 & 2.8 & 3.0 \\
 &(3400 - 3600) & (4100 - 4700) & (1 - 2)  & & & (0.899 - 0.971) &(1 - 2)  & &   \\
HR Peg     & 3500  & 4900  & 0  & -0.34 (13)  & 0.17 & 0.500 & 0  & 2.2 & 2.3 \\
 & (3500 - 3600) & (4500 - 5200) & (0 - 4)  & & & (0.500 - 0.752)  & (0 - 1)  & & \\

\hline
\end{tabular}
\tablefoot{In the \teff, $\log g$, and [s/Fe] columns, the numbers between brackets denote the values spanned during the $\log g$ iterations, while in the C/O column they indicate the error on C/O from the spectral synthesis of the CH bands. In the $L$ column, they indicate the luminosity error due to the  error on the Gaia EDR3 parallax.
The numbers in  brackets in the [Fe/H] column indicate the number of Fe lines used to derive [Fe/H] and the next column indicates the standard deviation (derived from the line-to-line scatter) on [Fe/H]. The current (M$_{\rm curr}$) and initial (M$_\mathrm{ini}$) masses have been derived from the locations of the stars in the HR diagram compared to STAREVOL tracks \shreeya{(Sect.~\ref{sectHRD})}.}
\end{table*}

%====================================

\section{Chemical abundance determination}\label{adundance determination}

The abundance determination methodology is the same as the one adopted in S18 and S19. We compared the observed spectra with synthetic spectra generated using Turbospectrum v15.1 \citep{turbospectrum} and  MARCS model atmospheres of S stars with the parameters derived in Sect.~\ref{parametersdetermination}. We have made use of the same input molecular line lists as in SVE17, atomic line list as in the Gaia-ESO survey \citep{GES2020} and varied the abundances till a satisfactory agreement could be found.
\subsection{Li}
\begin{figure}
    \centering
    \includegraphics[scale=0.4,width=9.5cm]{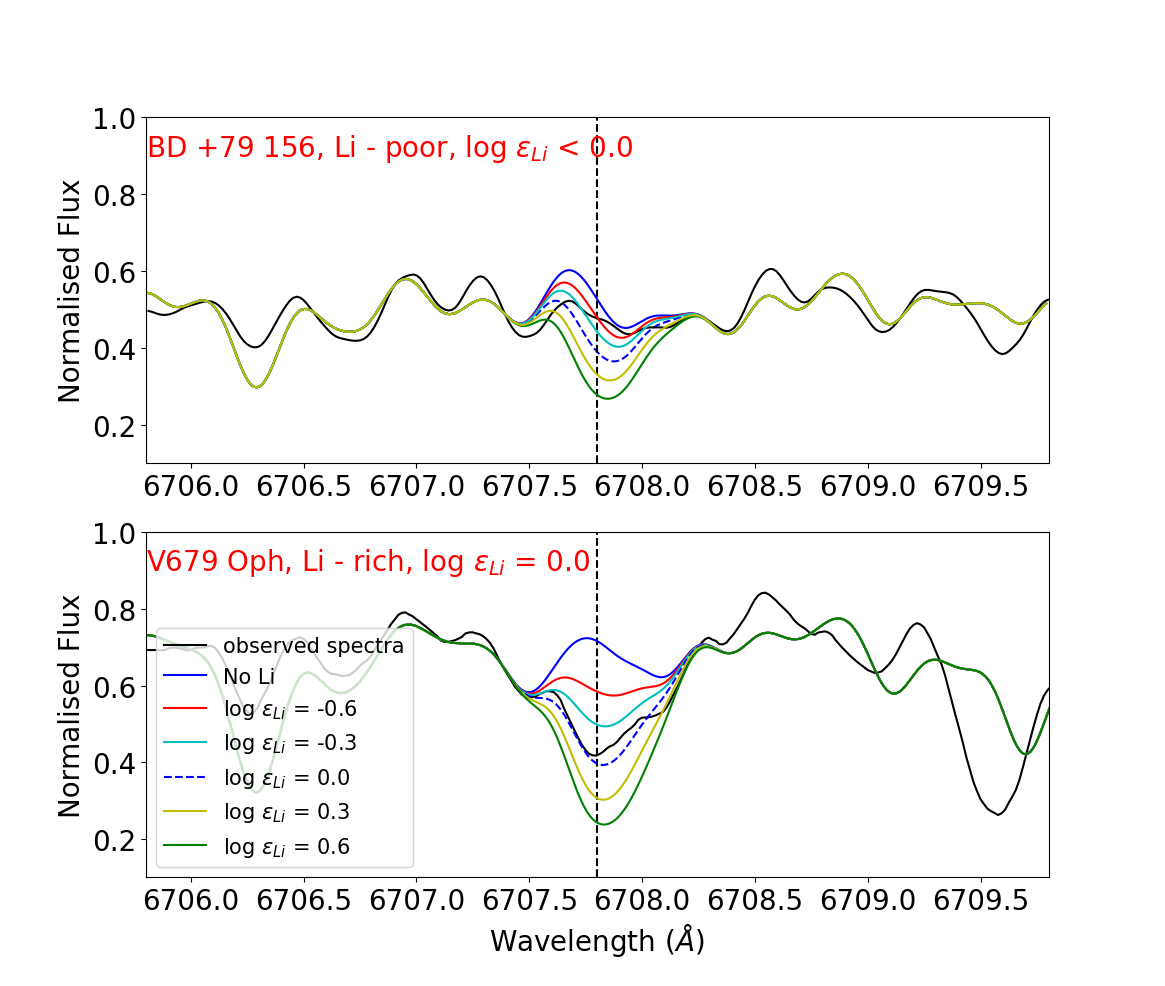}
    \caption{Illustration of the fit quality between the observed and synthetic spectra of BD~+79\deg156 (top panel) and V679~Oph (bottom panel) around the Li line at 6707.8 \AA.}
    \label{fig:Liexample}
\end{figure}

We used the \ion{Li}{i} line at 6707.8 \AA\ to derive the Li abundance. 
This line is known to be blended with a neighboring \ion{Ce}{ii} line at 6708.099~\AA\ in the warmer post-AGB stars \citep{reyniers2002}. However, we did not identify any dominating blend from this line to the \ion{Li}{i} line in our sample stars.
In Fig.~\ref{fig:Liexample} we present examples of stars with a good (bottom panel) as well as bad (top panel) 
spectral-synthesis fit of the  considered Li line.
We could derive the Li abundance for only three stars from our sample, namely HR~Peg, V679~Oph and Vy~12. For these stars, the absorption features of the Li doublet located at 6707.8~\AA~are very clear. For the rest of the sample, there were severe blends in the Li line (top panel of Fig.~\ref{fig:Liexample}), so only upper limits on the Li abundances could be derived. The results are listed in Table ~\ref{tab:Liabund}. 

The Li abundances of all the intrinsic S stars of our sample are generally low except for HR~Peg. HR~Peg was found to be a Li-rich star also by \cite{vanture2007} who classified the star as a high-mass star ($M > 3$~\msun) with hot bottom burning (HBB) as an explanation for the Li abundance. However, our mass estimate for HR~Peg is $M_\mathrm{curr}$ = 2.2~\msun{}, in agreement with the 2.0~\msun{} value found by \citet{pieter2015}.
Therefore, HR~Peg does not appear to be massive enough to be producing Li through HBB. Other plausible explanations involve some extra mixing in low-mass AGB stars \citep[e.g.][]{charbonnel2000, uttenthaler2007,uttenthaler2010} or the engulfment of planets/brown dwarfs \citep{Siess99}.

\subsection{C, N, O} \label{CNO}
We used the CH bands around 4250~\AA~to determine the C abundance. The sensitivity of these bands to the carbon abundance is limited in S-star spectra because CH bands 
are blended with mainly TiO and ZrO (as can be seen from Fig.~16 of SVE17). 
It was not possible to derive the O abundance because the $\lambda$~6300.3~\AA\  \ion{O}{i} line lies in a severely blended region. We used the solar oxygen abundance \citep{asplund2009} scaled with respect to the metallicity and included an alpha enhancement ([$\alpha$/Fe] =~$-0.4\;\times$ [Fe/H]) for [Fe/H] $\ge -0.5$.
The uncertainty on C/O was estimated from the range of values of the C/O providing an acceptable fit to the CH G-band.

The N abundance was then determined from the CN lines available in the region 8000-8100~\AA. The line list for these CN lines was taken from \cite{sneden2014}.

\subsection{[Fe/H]}
We used the same line lists as S18 and S19 (see also Appendix Table~\ref{linelist}) for Fe line synthesis. 
Between 9 and 15 Fe lines were used to derive the metallicity for all the sample stars, except for 
the star Vy~12, for which  only six Fe lines could be used, as the other lines were strongly blended by molecules because of the high C/O of this object (0.97). The derived metallicity as well as the standard deviation due to line-to-line scatter are listed in Table~\ref{finalparams}. 

\subsection{Light s-process elements (Sr, Y,  Zr, Nb)}
\begin{figure*}
    \centering
    \includegraphics[scale=0.4]{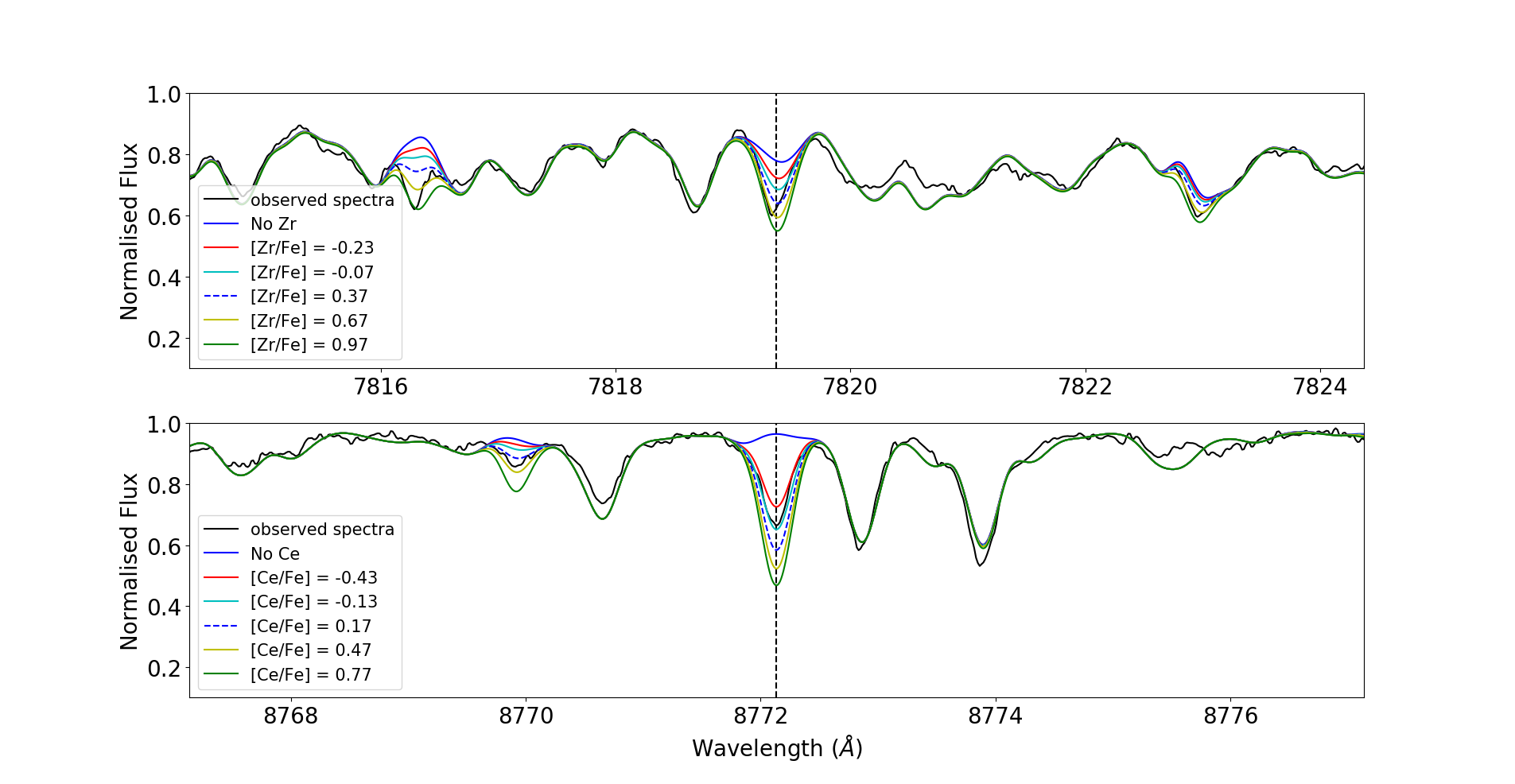}
    \caption{Illustration of the %quality of match 
    fit between the observed and synthetic spectra for the S star HR Peg. The top panel presents 5 \AA~around the \ion{Zr}{I} line at 7819.37 \AA~and the bottom panel around the \ion{Ce}{II} line at 8772.135 \AA.}
    \label{spectralfitexample}
\end{figure*}

The line list of S18 was used and complemented as documented in Appendix Table~\ref{linelist}. 
The strontium abundance could be derived only for HD~288833, BD-18\deg2608, and V679~Oph. The Y abundance was measured for all the sample stars. 
The Zr abundance was measured for all stars except KR~CMa using the two \ion{Zr}{i} lines at 7819.37 and 7849.37~\AA~with transition probabilities from laboratory measurements \citep{Zrlines}. 
Some \ion{Nb}{i} lines from Appendix Table~\ref{linelist} could not be used in some stars. In fact, none of the \ion{Nb}{i} lines could be used for CSS~151 because of the low S/N of its spectrum. The Nb abundance could therefore not be derived for this star.

\subsection{Heavy s-process elements (Ba, Ce, Nd)} 
We derived the barium abundance using only one \ion{Ba}{i} line located at 7488.07~\AA. The cerium abundance was determined using the \ion{Ce}{ii} lines from S18 together with some new lines located at 7580.91, 8025.57, 8394.51, 8405.25, and 8769.91~\AA. There is a huge scatter in the Ce abundances for HR Peg, $\omicron^1$ Ori, KR~CMa, BD+79\deg156, and HD~64147 despite the use of \ion{Ce}{ii} lines displaying satisfactory fits. The bottom panel of Fig.~\ref{spectralfitexample} illustrates the fact that even neighboring \ion{Ce}{ii} lines (at 8769.91  and 8772.13~\AA) can provide discrepant abundances (by 
$\sim$~0.4 dex in this case). As explained in \cite{Drisya}, cerium abundances derived using only 
\ion{Ce}{ii} lines above 7000~\AA~are $\sim$~0.3 dex lower than the ones derived from \ion{Ce}{ii} lines in the range 4300-6500~\AA. These authors mention that this difference might be imputable to 
non-LTE (non-local thermodynamic equilibrium) effects. This could be the reason for the sub-solar cerium abundances in some of our stars, as they were determined using only red \ion{Ce}{ii} lines ($\lambda > 7000$~\AA). These uncertain abundances are marked with a colon in  Tables~\ref{abundance1} and \ref{abundance2}.

\subsection{Other heavy elements (Pr, Sm, Eu)}
Good \ion{Eu}{ii} and \ion{Pr}{ii} lines are located in the bluer part of the spectrum and could be used in some stars where the blending was weak. The Sm abundance was derived when possible using 
the \ion{Sm}{ii} 7042.20 and 7051.55~\AA~lines. These lines are well fitted and yield consistent abundances.

\subsection{Tc}
All three \ion{Tc}{i} resonance lines at 4238.19, 4262.27 and 4297.06~\AA~are heavily blended \citep{little1979}. We first derived the other s-process abundances in order to reproduce these blends as precisely as possible. The 4262.27~\AA~line is the best reproduced by our spectral synthesis. Its blends, consisting in two neighbouring lines of \ion{Nb}{i} (at 4262.050~\AA) and \ion{Gd}{ii} (at 4262.087~\AA) 
were identified in  \cite{sophie1999}. The Tc abundance was derived from the 4262.27~\AA~\ion{Tc}{i} line in our sample stars as well as in all intrinsic S stars from S18 and S19 (Table~\ref{tab:Liabund}).

\subsection{Uncertainties on the abundances}\label{abundanceuncertainty}

The uncertainties on the abundances have been computed using the ones of the S star V915~Aql investigated in S18. The atmospheric parameters of V915~Aql are representative of those of most of our sample stars, hence we computed the elemental abundance error by quadratically adding the elemental standard deviation due to line-to-line scatter, the abundance errors due to parameter uncertainties of V915~Aql as derived in S18 (see also Appendix~\ref{ErrorTc_Li}), and a term of 0.1~dex to take into account continuum placement errors. For abundances that were derived using only one line, an arbitrary line-to-line scatter of 0.1~dex was assumed. Error bars on Tc abundances were estimated as the range of Tc abundances providing an acceptable fit of the \ion{Tc}{i} 4262.27~\AA\ line. The final elemental uncertainties on abundances are listed in Appendix Tables~\ref{abundance1}, \ref{abundance2}, and \ref{abundance3}.

\section{HR diagram of S stars} \label{sectHRD}
\begin{figure*}[hbt!]
\begin{centering}
    \mbox{\includegraphics[scale=0.4,trim={1.5cm 0cm 0cm 0cm}]{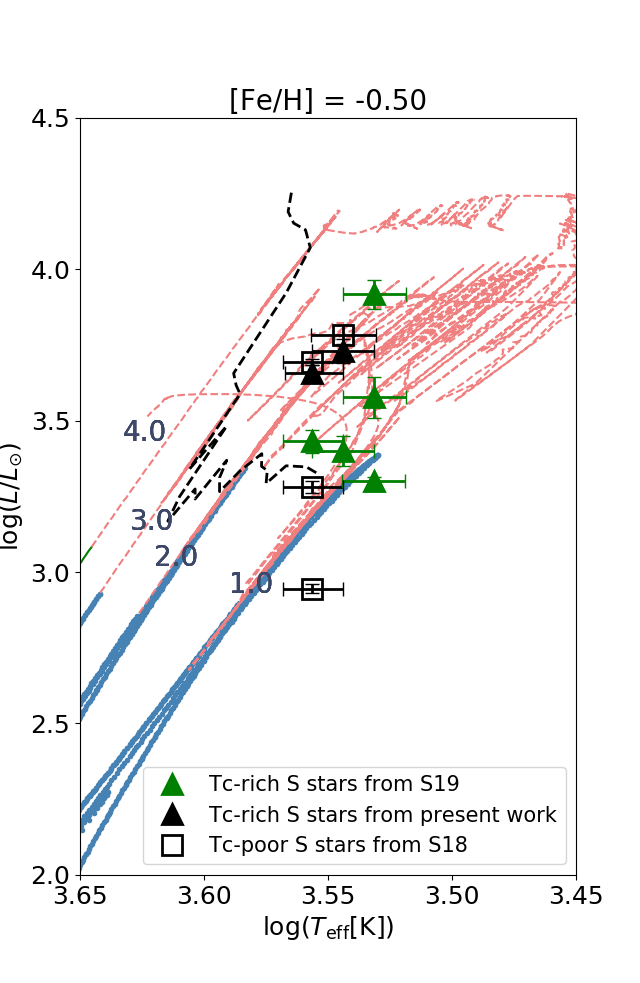}}   
    \hspace{0px}
    \mbox{\includegraphics[scale=0.4,trim={1.5cm 0cm 0cm 0cm}]{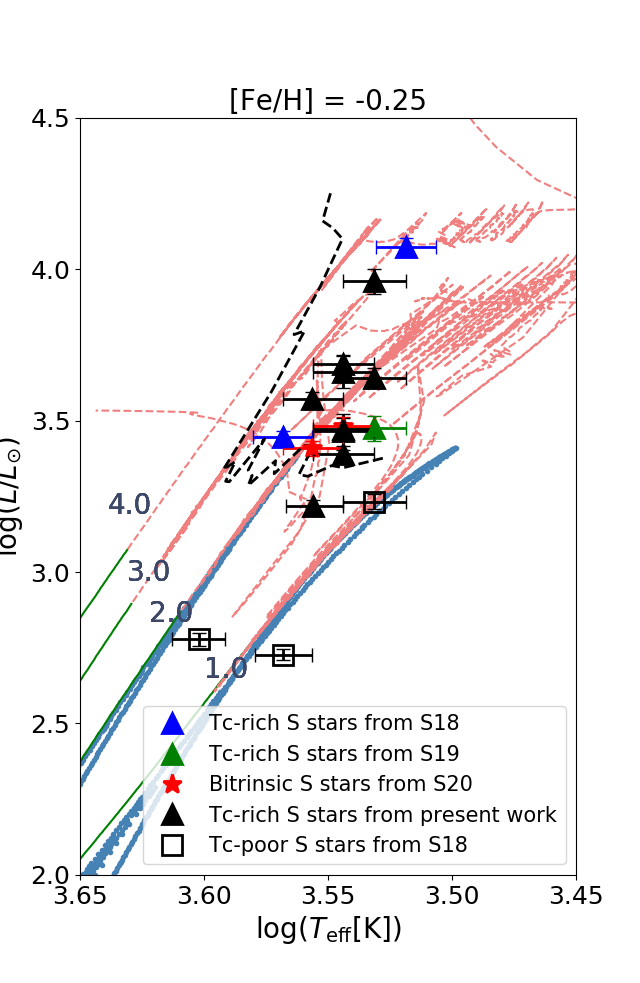}}   
    \hspace{0px}
     \mbox{\includegraphics[scale=0.4,trim={1.5cm 0cm 0cm 0cm}]{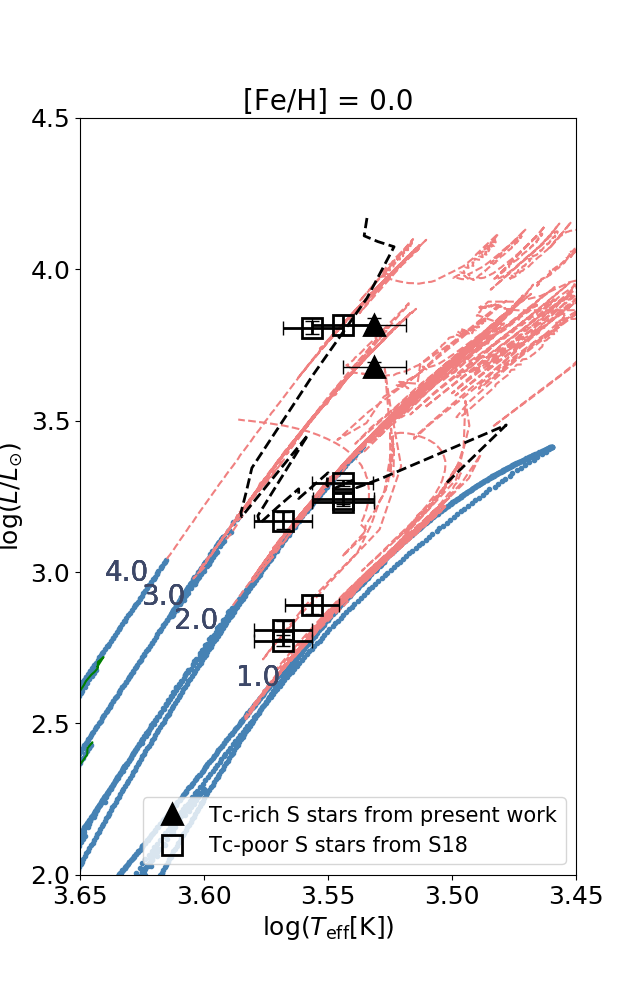}}   
    
    \caption{\label{HRD}  HR diagram of intrinsic (filled triangles) S stars from our large sample and extrinsic (open squares) S stars from S18 along with the STAREVOL evolutionary tracks corresponding to the closest metallicities. The red giant branch is represented in blue, the core He-burning phase in green, whereas the red dashed lines correspond to the AGB tracks. The black dashed line represents the predicted onset of the third dredge-up, i.e., the lowest stellar luminosity following the first occurrence of a TDU episode \shreeya{(down to 1.5, 1.3, 1.0 \msun\ for [Fe/H] = 0.0, -0.25, -0.50 respectively).}%(when the convective envelope penetrates in the intershell zone where the thermal pulse was existing)
     }
\end{centering}
\end{figure*}

In Fig.~\ref{HRD} we present an HR diagram collecting
Tc-rich S stars studied in S18, S19, and S20 as well as those of the present work, together with Tc-poor stars from S18. \shreeya{We used the final stellar parameters presented in Table~\ref{basic data}. The errors on \teff~are taken from Table~\ref{basic data}.
For HD~288833, V812~Oph, AA~Cam, and V679~Oph for which \teff~did not change during the \logg\ iterations, we imposed a standard symmetric error of 100~K on their \teff. The asymmetric error on the luminosity was derived after propagating the error on the parallax.} The evolutionary tracks displayed in Fig.~\ref{HRD} were computed with the STAREVOL code \citep{Siess2000} and are described in detail in \cite{escorza2017}. Briefly, we use standard input physics with a mixing-length parameter $\alpha=1.75$, grey surface boundary conditions and the \cite{asplund2009} solar mixture. Opacity enhancement due to the formation of molecules in carbon-rich atmospheres is also accounted for following the formulation of \cite{Marigo2002}. The \cite{schroder} mass-loss prescription is activated up to the end of core helium burning followed by the \cite{vw93} formulation during the AGB phase. We also consider overshooting below the convective envelope following the exponential decay expression of \cite{herwig97} with the parameter $f_\mathrm{over} = 0.01$. Models were computed for various metallicities, including [Fe/H]~=~0, -0.25 and -0.5. 

There is a clear segregation between the location of intrinsic and extrinsic S stars in the HR diagram. 
Indeed, whatever the considered metallicity range, intrinsic S stars are always located above the black dotted line (see Fig.~\ref{HRD}) marking the predicted onset of TDU for masses above 1.5~\msun. Whereas it is obvious that Tc-rich stars must be TP-AGB stars (because of the Tc detection), such a consistency between the luminosity of the third dredge-up onset in theoretical models, and the luminosity of observed intrinsic S stars, which are the least evolved objects identified as TP-AGB stars, was not as clearly demonstrated so far.

Table~\ref{lowlumlimit} summarizes the 
observational constraints on the TDU first occurrence in terms of luminosity, initial mass and metallicity.
For each mass and metallicity bins, we indicate the lowest luminosity of intrinsic S stars from our so-called {\em large sample} collecting stars from  S18, S19, S20 and from the current work. 
These luminosities represent an observational upper limit on the TDU first occurrence in AGB stars, in the sense that Tc-rich stars are observed at such luminosities, so that TDU must have already occurred by the time the star reaches this position in the HR diagram. However, since S stars are the first objects on the AGB to show clear signatures of TDU events (Tc and ZrO), this upper limit must be quite close to the genuine TDU occurrence line.
It can be used as an observational constraint to be satisfied by the stellar evolutionary models of the corresponding masses and metallicities.

\begin{table*}
\caption{Observed lowest luminosities of intrinsic S stars in different mass and metallicity bins (collected from this work, S18, S19 and S20). }
    \label{lowlumlimit}
    \centering
    \begin{tabular}{c|l|c|c|c}
    \hline
    \hline
          & Stars & Initial mass (\msun) & [Fe/H] & $L_{\rm TDU}$ ($L_{\odot}$)  \\
         \hline
         M$_\mathrm{ini}$  < 1.0 \msun  & V915 Aql & 1.0 & -0.5 & 2000 \\
         %                             & HD 357941 & 0.9 & -0.27 & 1357 \\
         \hline
         1.0 \msun < M$_\mathrm{ini}$ < 1.5 \msun &  IRAS 06000+1023 (CSS 182) & 1.3 & -0.40 & 2500 \\
         & HD 288833 & 1.4 & -0.30 & 1700\\
         
         \hline
         1.5 \msun < M$_\mathrm{ini}$ < 2.0 \msun & BD -18 2608 & 1.6 & -0.31 & 2400 \\
          & V812 Oph & 2.0 & -0.37 & 3000 \\
          \hline
         2.0 \msun < M$_\mathrm{ini}$ < 2.5 \msun & BD+79\deg156 & 2.1 & -0.16 & 2600 \\
         & V679 Oph & 2.1 & -0.52 & 4600 \\
                                                  %& V812 Oph & 2.5 & -0.01 & 4200 \\
                                                 
         \hline
         2.5 \msun < M$_\mathrm{ini}$ < 3.0 \msun & HD 64147 & 2.9 & -0.67 & 5400 \\
         & CSS 1152 & 3.0 & -0.14 & 4400 \\
         & KR CMa & 3.0 & -0.01 & 4800 \\
         \hline
         M$_\mathrm{ini}$ > 3.0 \msun   & CSS 151 & 3.2 & -0.25 & 4600 \\
                                        & V1139 Tau & 3.5 & -0.06 & 6600 \\
         \hline
         \hline
    \end{tabular}
\end{table*}

\section{Discussion on the abundances and comparison with STAREVOL nucleosynthesis predictions}
\label{sectabundances}

\begin{figure*}
    \centering
    \includegraphics[scale=0.8,trim={0cm 0cm 0cm 0cm}]{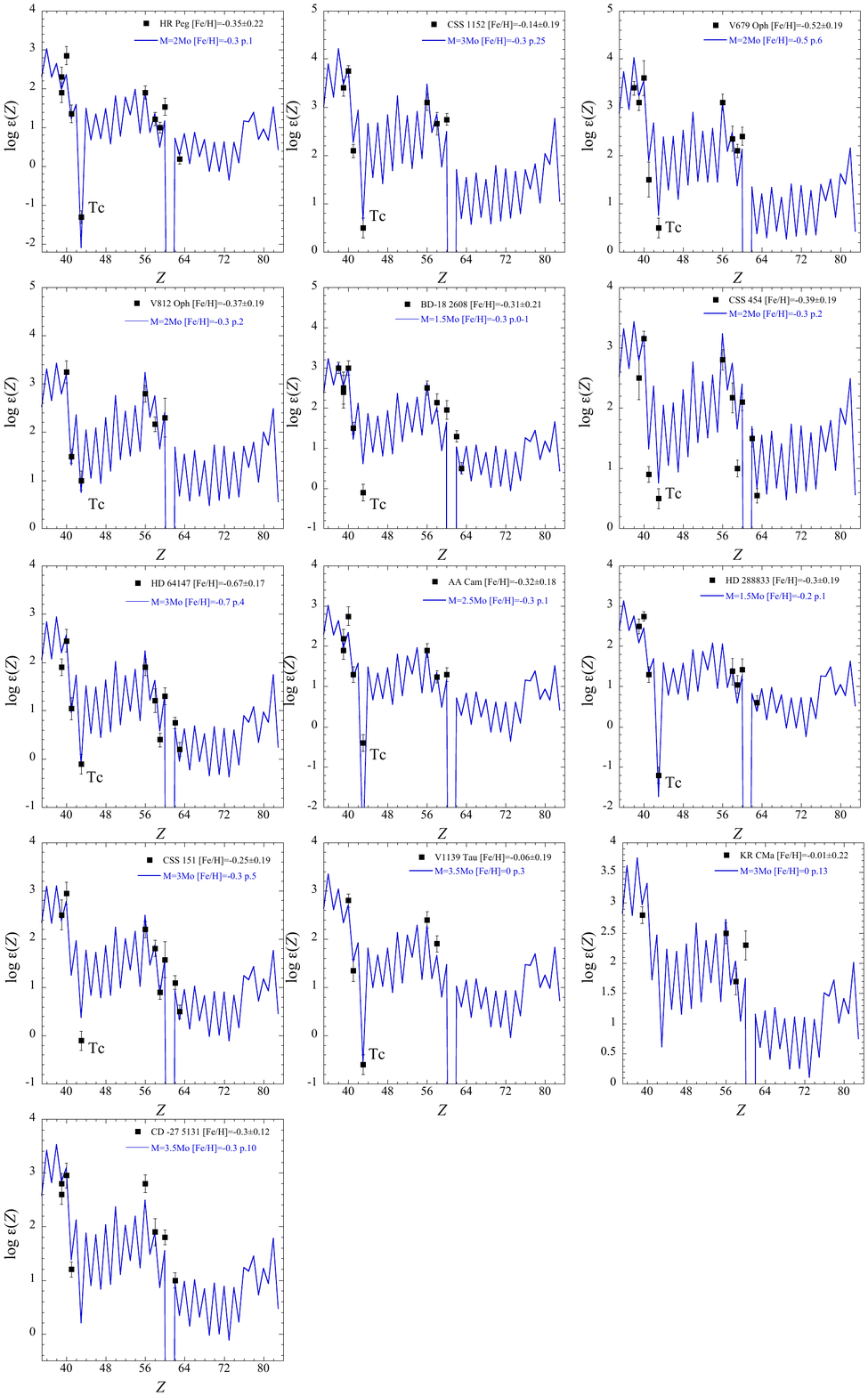}
    \caption{Measured s-process element abundances compared with the nucleosynthesis predictions. The blue line represents the nucleosynthesis predictions compatible with the mass and metallicity of the individual stars. The number of pulses $n$ required to best-match the measured abundances is mentioned in the label of every panel as "p.$n$".}
    \label{fig:my_label}
\end{figure*}

The elemental abundances derived following the methodology presented in Sect.~\ref{adundance determination} allow us to investigate nucleosynthesis in intrinsic S stars. We compare the measured abundance profiles of our sample stars with the nucleosynthesis calculations of the STAREVOL code \citep{goriely&siess} which uses an extended network of 411 species. 

The predicted and measured abundance distributions are presented in Fig.~\ref{fig:my_label}. 
The pulse number is chosen in order to optimally match both the overabundance level and the  location  in  the  HR  diagram. We find a good overall agreement between the predicted and measured distributions of heavy elements. In particular,  the peak of heavy s-elements is well reproduced in all stars. 
However, problems persist with the light elements such as carbon or oxygen, as discussed in Sect. 7.1, and
with some heavy elements, e.g. Ce as discussed in Sect. 5.4. The models do account for most of the derived Tc abundances; however, we note that abundance predictions for Tc are extremely sensitive to the pulse number and to the amount of dilution in the stellar envelope due to the initial absence of Tc in the star. Hence, the agreement between predicted and measured abundances may be poor in some cases, for e.g., CSS~151, BD~-18\deg2608, and CSS~454.
We now investigate specific element ratios.

\subsection{[C/Fe] and [s/Fe]}\label{Sect:coverfeVssoverFe}
\begin{figure}[hbt!]
\mbox{\includegraphics[scale=0.43]{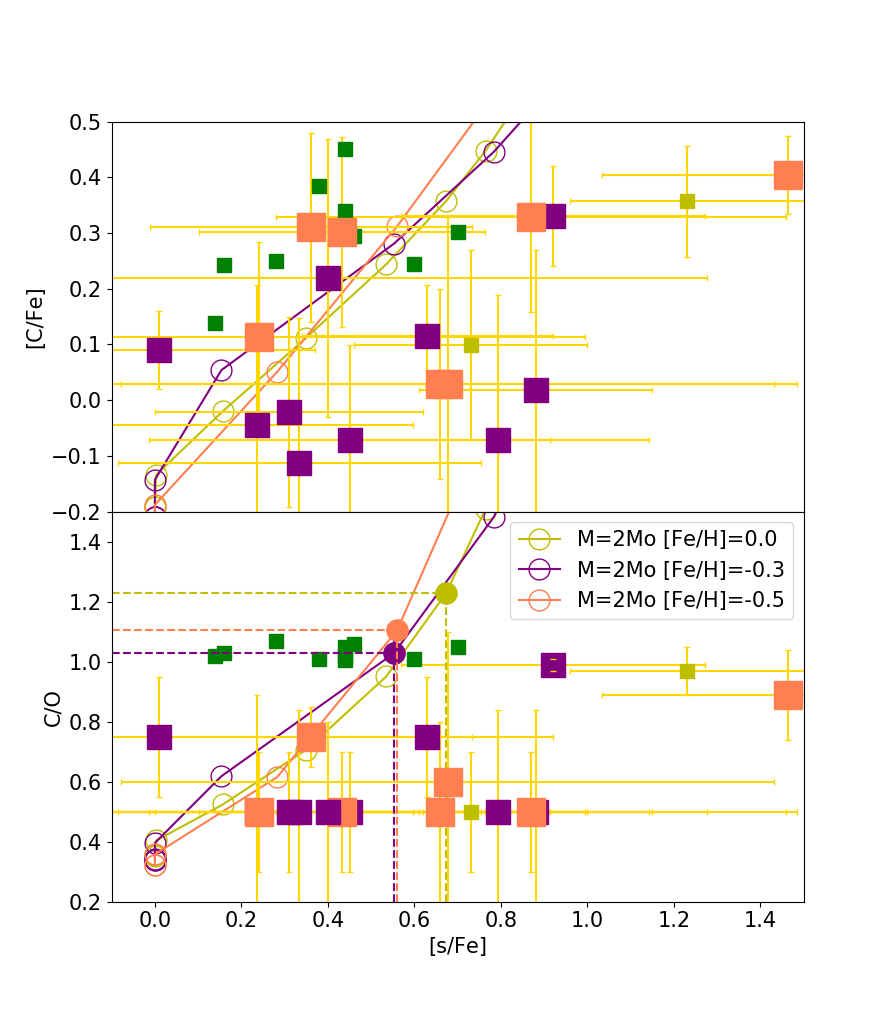}}  

    \caption[]{
    Top panel: measured [C/Fe] abundances as a function of [s/Fe] for our large sample of Tc-rich S stars (light green, purple and orange squares)
    where the size of the 
    symbol increases with decreasing metallicity, considering 3 metallicity bins: 
    [Fe/H] in [0.0;$-0.2$]: light green,
    [$-0.2$; $-0.4$]: purple, below $-0.4$: orange.
    Tc-rich carbon stars from \cite{abia2002}
     are shown as green squares.
    Bottom panel: derived C/O ratio versus [s/Fe] for the same stars.
    Predictions for a 2~\msun~ STAREVOL model at metallicity
    [Fe/H] = 0.0 (light green), $-0.3$ (violet) and $-0.5$ (orange) are overplotted.
    The empty circles along the tracks indicate the successive TDUs, while the three filled circles mark the first TDU allowing to reach C/O$>1$ in each model.}
    \label{carbonvssindex}

\end{figure}

The surface composition of TP-AGB stars should reflect the addition of $^{12}$C originating from the He-burning shell and s-process material produced either radiatively in the interpulse or in the convective thermal pulses for more massive objects. The TDU is then responsible for transporting these products to the stellar surface. As the star evolves on the TP-AGB, its carbon abundance should thus increase along with its s-process over-abundances. Eventually the star becomes carbon-rich  when the C/O ratio  exceeds unity. 
However, massive AGB stars with M$_\mathrm{ini} \ga 4$~\msun\  experience HBB and can efficiently burn the dredged-up carbon to produce mainly $^{14}$N. In the top panel of Fig.~\ref{carbonvssindex} we present 
the carbon abundance as a function of s-process abundance for our intrinsic S star sample. 
The [s/Fe] index has been calculated using the Y, Zr and Ba abundances and is listed in Table~\ref{hslsandsindex}. In order to compare this trend with that of the stars from the next evolutionary stage, we added in Fig.~\ref{carbonvssindex} the Tc-rich carbon stars from \cite{abia2002}.
A larger marker size indicates a lower metallicity 
(stars are grouped in three [Fe/H] bins: [$-\infty, -0.4$], [$-0.4 , -0.2$], and [$-0.2, 0$]).

In the top panel of Fig.~\ref{carbonvssindex}, the trend of increasing [C/Fe] with increasing [s/Fe], expected for TP-AGB stars, is not marked. Nevertheless, the S stars most enriched in s-process elements are the ones with the highest carbon abundance. The carbon stars have higher carbon abundances than most S stars
but are characterized by [s/Fe] indices in the same range as those of the bulk of S stars. A group of intrinsic S stars have similar [s/Fe] and [C/Fe] ratios as carbon stars.
This can be explained as follows: the carbon stars from \cite{abia2002} have solar metallicities while the intrinsic S stars belonging to the group with similar [C/Fe] (in the range 0.2 to 0.35 dex) have lower metallicities (in the range -0.2 to -0.5 dex). Because [$\alpha$/Fe] increases with decreasing metallicity in the range $-0.5 < $[Fe/H]$< 0$, this group of S stars has a higher initial [O/Fe] compared to carbon stars, enabling their O-rich classification. 

The [C/Fe] abundances in this work were derived from the C/O ratios assuming solar oxygen abundance scaled with respect to metallicity and taking into account an $\alpha$-element enhancement (Sect.~\ref{CNO}). 
The determination of [C/Fe] is therefore quite indirect.
We thus present in the bottom panel of Fig.~\ref{carbonvssindex} the C/O ratios. 
Their determination is very robust, given the high sensitivity of molecular bands to the C/O ratio. 
As expected, there is no overlap between C/O ratio of C stars and S-type stars.
We note that 3 S-type stars have C/O ratios close to, but definitely lower than 1: indeed their C/O of 0.899, 0.971, and 0.998, produces an observed spectrum markedly different from the spectrum of an SC star (characterized by C/O $=1$).

A common misunderstanding is that S stars have C/O $=1$. This statement is often encountered in the literature even as a definition of S-type stars. Once again \citep[see also][]{SVE17}, we stress here that only S stars with the highest [s/Fe] values have C/O ratios approaching (but not reaching) unity.
Most S-type stars have intermediate C/O ratios (between 0.5 and 0.8). Stars with C/O $=1$ are actually classified as SC or CS stars.

Figure~\ref{carbonvssindex} also compares the measured carbon and s-process abundances with %those from
the nucleosynthesis predictions at three different metallicities 
([Fe/H] = 0.0, -0.3 and -0.5). The filled circles along the tracks mark the thermal pulses turning the C/O ratio to values above 1, thus changing the (model) stars into carbon stars. First, the different models predict a tight correlation between C/O and [s/Fe], whereas there is a large scatter in measured [s/Fe] at a given C/O. Second, theoretical calculations show a much faster increase of C/O with [s/Fe] than what is actually measured in stars. For example, all models predict that stars with [s/Fe] $\ge 0.55$ should have C/O $>1$ (and be carbon stars), whereas many S stars (which must have C/O $< 1$) are observed with such large s-process enrichments. In other words, in theory, high s-process enrichments go along with very high C/O ratios which are incompatible with S-star classification. The reason why carbon stars do not show such large s-process overabundances (stronger line blending? dust obscuration of the most evolved objects?) is yet unclear.

Finally, we remark that Fig.~\ref{carbonvssindex} reveals the existence of one S star, \omiori, with very mild -- if any -- enrichment in s-process elements (see [s/Fe] in Table~\ref{hslsandsindex}, Figure~4 and Table~A.1 from S20 for individual heavy-element abundances). This star is  another example of the "Stephenson M-type stars" uncovered by \citet{smithlambertfeb1990}. Nevertheless, it shows clear Tc signatures (Figure~1 of S20). The possible reasons for the low [s/Fe] index of \omiori\ are discussed in detail in S20 and \cite{jorissen2019}.

\begin{figure}[hbt]
    \centering
    \includegraphics[scale=0.37]{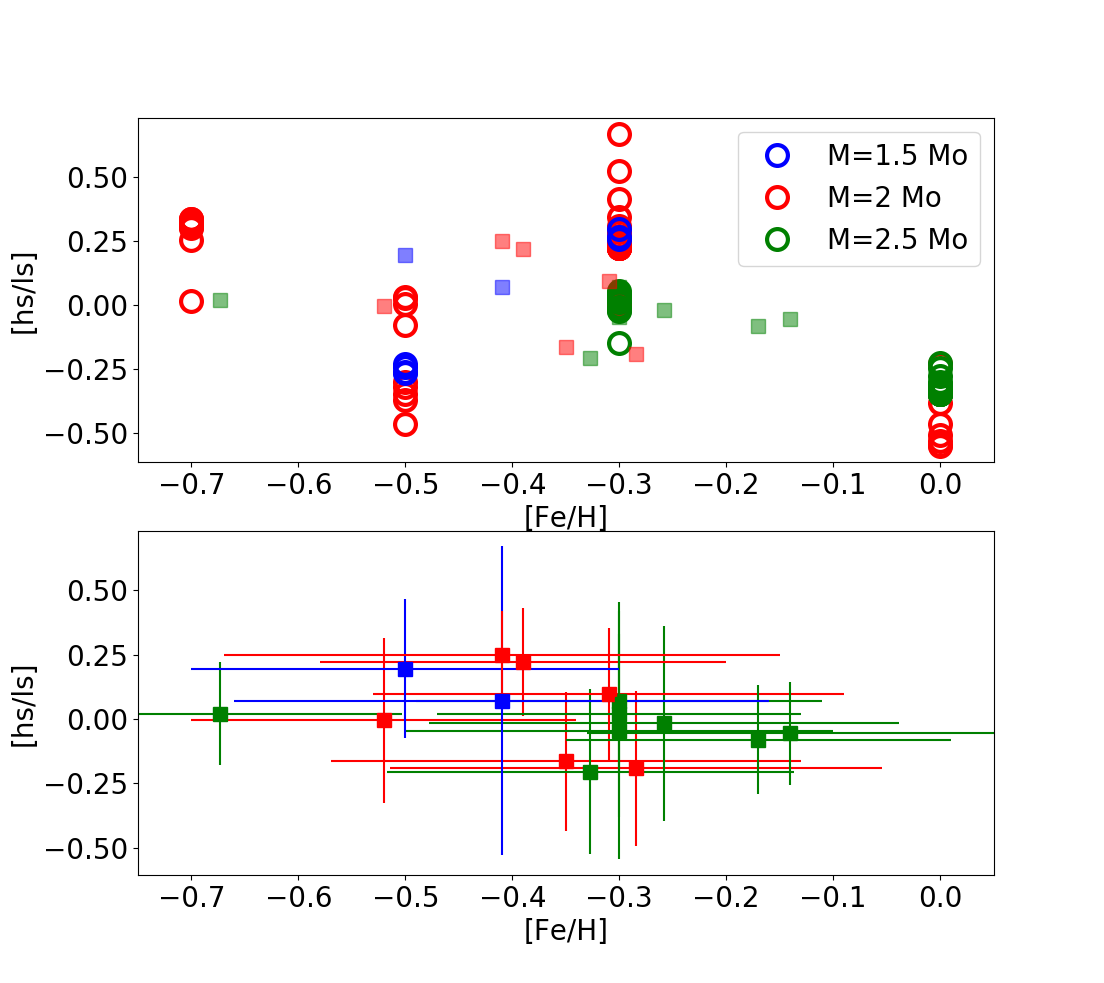}
    \caption{The distribution of [hs/ls] as a function of metallicity for the intrinsic S stars of our study (filled squares). The hs-index has been calculated using Ba and Nd, and the ls-index using Y and Zr. The symbols are color-coded with respect to the initial mass in the bins M$_\mathrm{ini}$ < 1.5~\msun~ (blue), 1.5 $\leq$ M$_\mathrm{ini}$ < 2.5~\msun~(red), and M$_\mathrm{ini}$ $\geq$ 2.5 \msun~(green). Note that these mass bins are designed to match the [hs/ls] STAREVOL predictions only available for M$_\mathrm{ini}$ = 1.5, 2, 2.5~\msun. Open circles denote the predictions from pulse to pulse for the different initial masses.}
    \label{FeoverHwithmeta}
\end{figure}

\subsection{[hs/ls] and metallicity}
We now investigate the potential correlations between the heavy (hs) to light (ls) s-process element ratio [hs/ls], mass and metallicity. The neutron irradiation index [hs/ls] is actually expected to increase with decreasing metallicity \citep{goriely-mowlavi-2000}, as the number of neutrons per iron seed increases.

From our measured abundances we find a large scatter in [hs/ls] with respect to metallicity (Fig.~\ref{FeoverHwithmeta}), within our limited range of metallicity.
~\cite{2015kenneth, 2016kenneth} also reported such a scatter in the [hs/ls] -- [Fe/H] plane from their study of s-process enriched post-AGB stars. However, the evolutionary link between post-AGB stars and intrinsic stars remains to be firmly established.
Figure~\ref{FeoverHwithmeta} also shows the absence of a clear correlation between the [hs/ls] ratio and the initial stellar masses.

When compared with the theoretical predictions
accounting for the pulse-to-pulse variation, plotted in the top panel of Fig.~\ref{FeoverHwithmeta}, 
we find that the overall range covered by our measured [hs/ls] indices is compatible with that of the 
predicted abundances of the theoretical models.
However, the [hs/ls] model predictions are not available for the complete metallicity range covered by our measured [Fe/H].

\subsection{Technetium abundances}\label{sect:tcabundance}

\subsubsection{[s/Fe], C/O and Tc}\label{s/FeVSTc}
\begin{figure}
    \centering
    \includegraphics[scale=0.4]{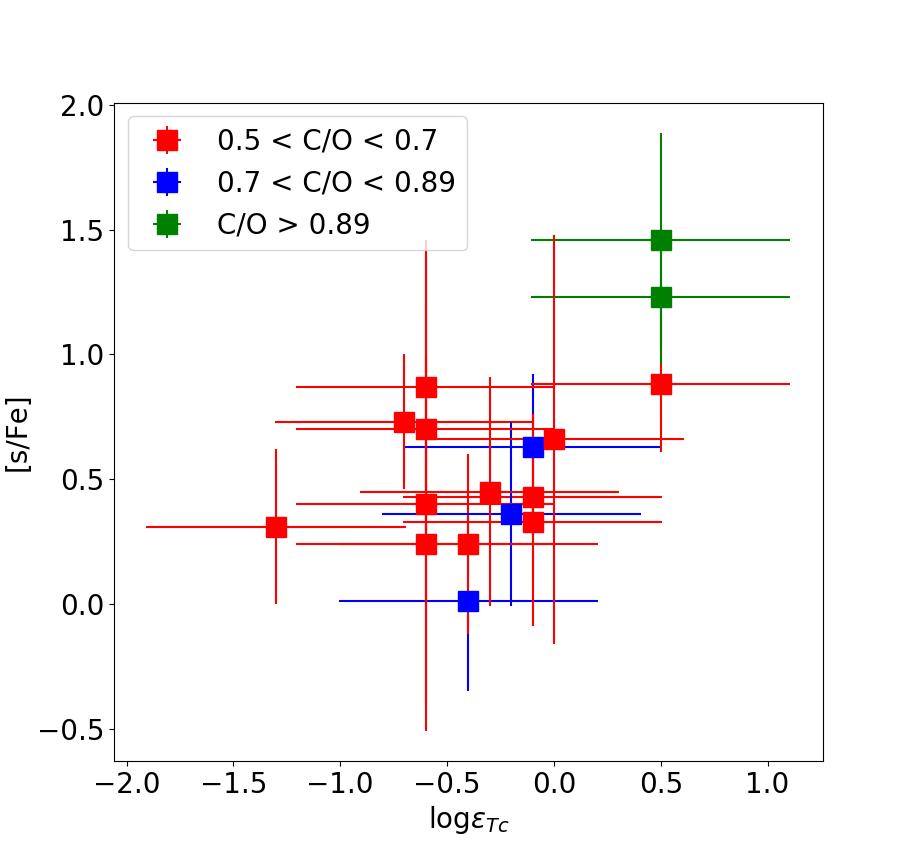}
    \caption{[s/Fe] index versus Tc abundance. The different colors represent different C/O ratio bins as described in the figure. }
    \label{fig:soerFeVsTc}
\end{figure}

Figure~\ref{fig:soerFeVsTc} presents [s/Fe] (calculated as described in Sect.~\ref{Sect:coverfeVssoverFe} \shreeya{using Y, Zr and Ba abundances}) as a function of the Tc abundance, for different C/O ratio bins.
The stars with the highest [s/Fe] and Tc abundances are also the ones with the highest C/O ratios, which is consistent with expectations from TP-AGB evolution. However, it is worth mentioning that whereas [s/Fe] is predicted to monotonically increase as the star ascends its AGB (see Fig.~\ref{carbonvssindex}, this is also true for C/O if there is no HBB, and for luminosity, set aside the luminosity variations during thermal pulses), the technetium abundance has a more complex behaviour, because the $^{99}$Tc half life (210 000 yrs) is not totally negligible with respect to the TP-AGB duration. For example, the evolution of the technetium-to-zirconium abundance ratio is not flat but shows a non-trivial evolution displayed in Fig.~2 of \citet{pieter2015}. The lack of a clear trend in Fig.~\ref{fig:soerFeVsTc} is therefore not surprising.

\subsubsection{Luminosity and Tc}\label{lumVSTc}
\begin{figure}
    \centering

    \includegraphics[scale=0.37]{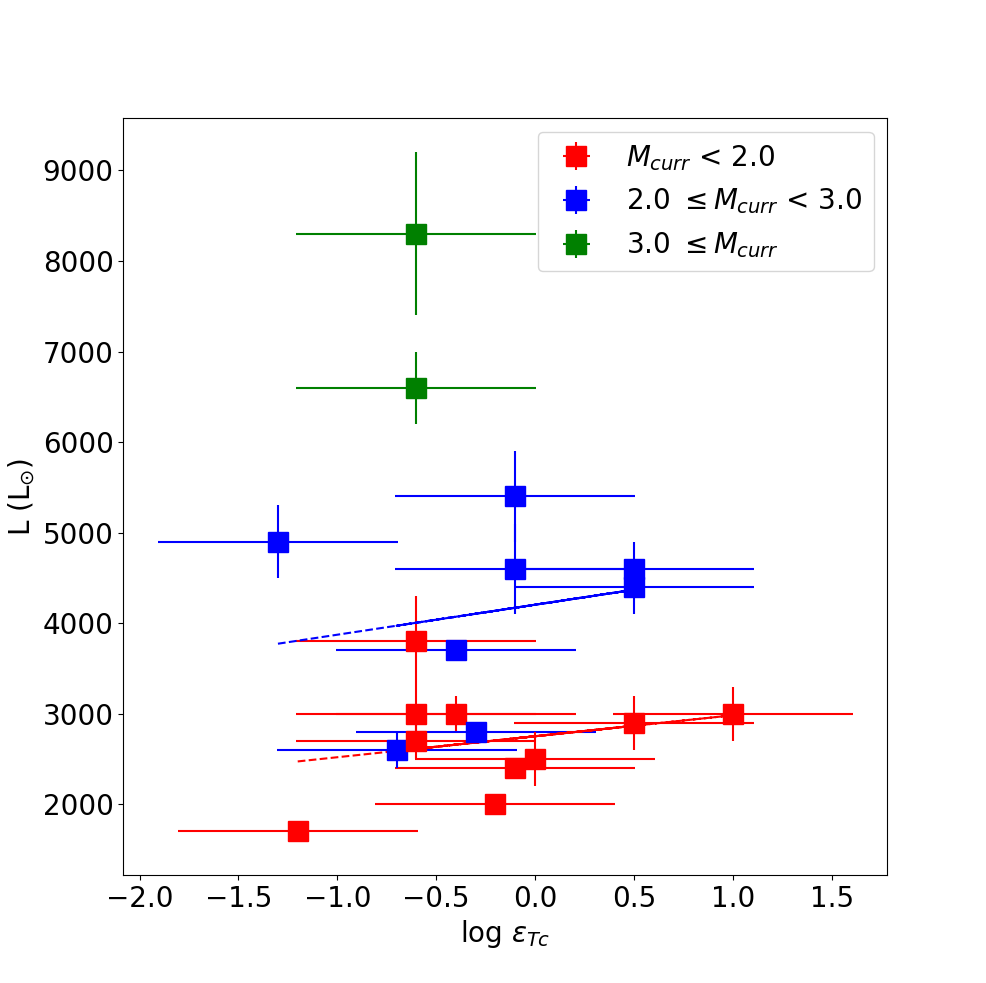}
    \caption{Stellar luminosity as a function of Tc abundance. The symbols are color coded according to their mass bin. The blue and red dashed lines represent the linear least-squares fit for the stars in the corresponding mass bin.   }
    \label{fig:lumvsTc}
\end{figure}

In Fig.~\ref{fig:lumvsTc} we compare the Tc abundances and the luminosities of the sample stars.
A large scatter in luminosity is present for each Tc abundance. 
If we focus on a specific mass bin, there is a loose trend of increasing Tc abundance with increasing luminosity, represented by the dashed least-square fit lines in Fig.~\ref{fig:lumvsTc}. We also find that the most luminous and massive star in Fig.~\ref{fig:lumvsTc} has a relatively moderate Tc overabundance. A likely explanation is the larger dilution of s-elements in the bigger envelope of more massive AGB stars. Besides, with increasing stellar mass, the mass of the $^{13}$C pocket responsible for the s-process, as well as that of the thermal pulses, are reduced because of the stronger compression of the shells induced by the larger core mass. Lower surface abundances of s-process elements are thus expected in higher-mass TP-AGB stars \citep{garcia2013}.

\subsection{The Zr-Nb plane and intrinsic/extrinsic stars segregation}

\begin{figure}
        \centering
        \includegraphics[scale=0.45]{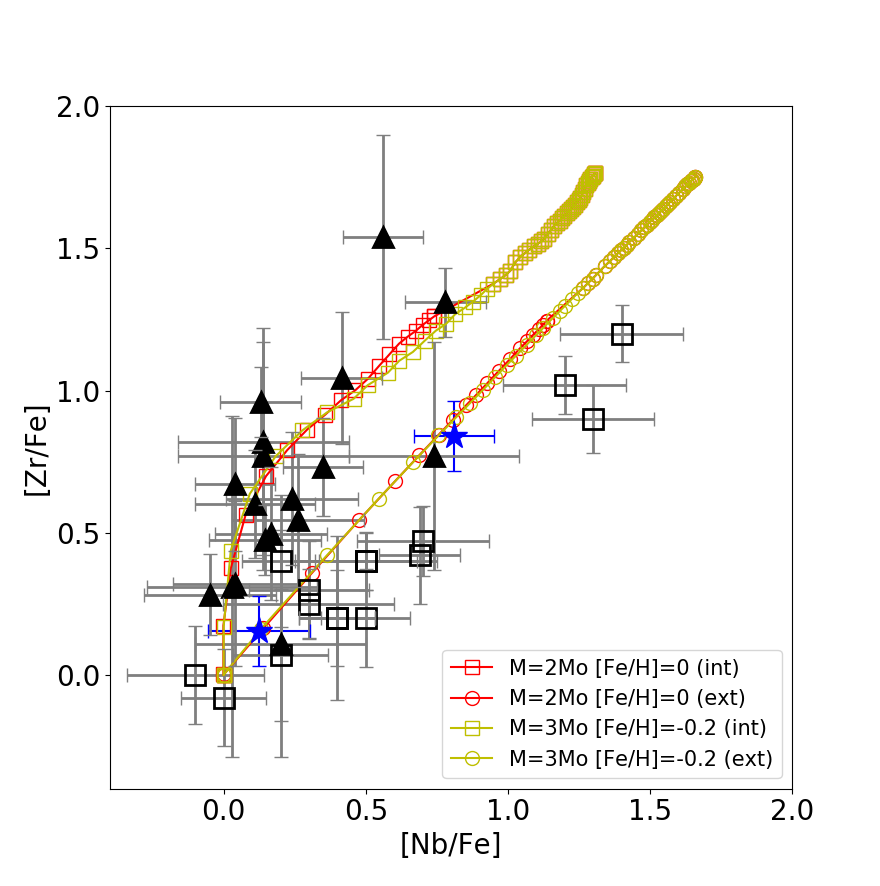}
        \caption{[Zr/Fe] as a function of [Nb/Fe] for the intrinsic S stars (black filled triangles) and extrinsic S stars (open squares) from S18 compared with the nucleosynthesis predictions for different initial masses and metallicities. The bitrinsic S stars from S20 are plotted with a blue star symbol. The "ext" and "int" labels next to the model parameters refer to extrinsic (respectively intrinsic) model abundances.}
        \label{ZrVsNb}
\end{figure}

A particular attention can also be paid to the Zr and Nb abundances that can be used as extrinsic/intrinsic star markers
\citep{pieter2015}. As already mentioned in Sect.~\ref{sec:intro}, niobium is mono-isotopic and can only be produced by the decay of~ $^{93}$Zr (with a half-life of $1.53\times 10^6$ yr). Intrinsic S stars are freshly producing s-process elements including $^{93}$Zr, and there is not enough time on the TP-AGB for  
a significant production of $^{93}$Nb. On the contrary, in extrinsic S stars, enough time has elapsed since the end of the nucleosynthesis in the companion for $^{93}$Zr to have totally decayed into $^{93}$Nb. 
As expected, intrinsic and extrinsic S stars follow different trends in the [Zr/Fe] -- [Nb/Fe] plane, where intrinsic S stars have [Zr/Nb] < 1 while extrinsic stars have [Zr/Nb] $\sim$ 1 \citep[][ S18]{pieter2015}. 
The Zr -- Nb plane has been studied extensively for extrinsically-enriched objects: 
the high Nb abundance in extrinsic stars has been demonstrated in various extrinsic-star families among which  CH and CEMP stars \citep{Drisya, Drisya2020}. The apparent shift of extrinsic stars from the
STAREVOL 2 – 3~\msun~ predictions by a few tenths of a dex is as well discussed in \cite{Drisya}. These authors speculated that this shift could arise due to the oscillator strengths of the two Zr lines used to derive the Zr abundance of the extrinsic S stars, which have a tendency to yield abundances that are about 0.1 dex too low in the benchmark stars of their sample (Arcturus and V762 Cas).
Here we populate the high-Zr, low-Nb region of this plane, adding constraints from the Tc-rich S stars.

Figure~\ref{ZrVsNb} confirms, on the basis of our so-called {\it large sample} of intrinsic S stars, the nice segregation between extrinsic and intrinsic stars in the [Zr/Fe] -- [Nb/Fe] plane (with one exception discussed below). We also compare the measured Zr and Nb abundances with nucleosynthesis predictions from STAREVOL  for intrinsic and extrinsic stars of 2 and 3 \msun, and metallicities [Fe/H] = 0.0 and -0.2. 
In Fig.~\ref{ZrVsNb}, the intrinsic S stars from our sample have [Zr/Fe] < 1.6 and follow the trend predicted for TP-AGB stars (open squares in red and green). It is worth mentioning that some $^{93}$Zr decay (inducing some $^{93}$Nb production) is both predicted and observed, as can be seen from the fact that highly-enriched (high-Zr) intrinsic stars, which are the most evolved S stars on the TP-AGB, tend to have the largest Nb enrichments, and nicely follow the STAREVOL inclined trend in the [Zr/Fe] -- [Nb/Fe] plane.

The two Tc-rich S stars \omiori~and BD+79\deg156 are 'bitrinsic' S stars as they are Tc-rich (intrinsic) 
but show signs of binarity together with a large Nb enrichment (for the corresponding [Zr/Fe] ratio), two evidences that they are also extrinsic stars, as discussed in S20. In addition to these two stars, the use of Gaia EDR3 parallaxes revealed a potentially new `bitrinsic' candidate, BD $+34^\circ$1698. We re-computed the s-process abundances of this star as its stellar parameters (mainly mass and \logg) changed when Gaia DR2 parallaxes were replaced by Gaia EDR3. The revised s-process abundances of BD$+34^\circ$1698 are presented in Appendix Table~\ref{abundtablelowmass}. It has a [Zr/Nb] ratio close to unity, along with clear signatures of Tc (see Fig.~C1 of S19), hence it qualifies as a `bitrinsic' candidate. 
\cite{wang} classified it as a candidate extrinsic S star based on its IRAS photometry. 
Two spectra taken with the Hermes spectrograph revealed a clear radial-velocity variation:  $V_r = 18.59 \pm 0.05$ km~s$^{-1}$ on JD $2\,457\,502.44$ and $V_r = 22.59 \pm 0.09$ km~s$^{-1}$ on JD $2\,459\,289.51$. Our detection of the binary motion associated to a clear Tc enrichment and a [Zr/Nb] ratio close to unity unambiguously classify BD $+34^\circ$1698 as a member of the restricted family of bitrinsic stars (S20).
Lastly, Fig.~\ref{ZrVsNb} confirms that the Zr-Nb analysis successfully serves as an additional test (apart from Tc) for the classification of S stars as intrinsic or extrinsic.

\section{Conclusions}

Thanks to the combination of Gaia EDR3 parallaxes and the high-resolution HERMES spectra, we have
determined the stellar parameters of a sample of 13 intrinsic S stars with metallicities in the range -0.7 < [Fe/H] < 0.
We then derived their s-process element abundances.
The heavy-element abundances of intrinsic S stars reveal their rich nucleosynthetic history. The main results from our study can be summarized as follows:\\
(i) The Gaia EDR3 HR diagram of S stars confirms that intrinsic S stars
are more evolved than their extrinsic counterparts.\\
(ii) The luminosity lower limits for the occurrence of TDU in different mass and metallicity ranges provided in Table \ref{lowlumlimit} can be used to constrain AGB evolutionary models.\\
(iii) The objects from our sample with the largest C/O ratios are also the ones with the largest [s/Fe], which is consistent with TP-AGB evolution predictions. \\
(iv) However, we clearly demonstrate the too rapid increase of the C/O ratio with respect to [s/Fe] in model predictions.\\
(v) The measured s-process abundances of intrinsic S stars are in good agreement with the AGB nucleosynthesis predictions for models of the corresponding mass and metallicity. In particular, the Zr and Nb abundances are matching very well the predicted trend for intrinsic S stars, confirming our previous finding (S18, S20) that the Nb abundance can be used as an intrinsic/extrinsic diagnostic as efficient as the Tc presence/absence. \\
(vi) We present Tc abundances for a large sample of intrinsic S stars (20 stars). We find that the stars with the highest C/O ratios tend to be the ones with the highest Tc abundances. \\

In conclusion, the current investigation of a sample of intrinsic S stars has extended our understanding of their properties, in particular their location in the HR diagram and their chemical characterization considering $\sim$ 10 chemical elements, including C/O ratio and the radioactive element technetium.

\begin{acknowledgements}
      The authors thank the anonymous referee for constructive comments.
      This research has been funded by the Belgian Science Policy Office under contract BR/143/A2/STARLAB. SS, SVE, SG, MG acknowledge support from the FWO \& FNRS Excellence of Science Programme (EOS-O022818F).
      SVE thanks {\it Fondation ULB} for its support. GW's research is supported by The Kennilworth Fund of the New York Community Trust. 
      Based on observations obtained with the HERMES spectrograph, which is supported by the Research Foundation - Flanders (FWO), Belgium, the Research Council of KU Leuven, Belgium, the \textit{Fonds National de la Recherche Scientifique} (F.R.S.-FNRS), Belgium, the Royal Observatory of Belgium, the \textit{Observatoire de Gen\`eve}, Switzerland and the \textit{Th\"{u}ringer Landessternwarte Tautenburg}, Germany. This work has made use of data from the European Space Agency (ESA)  mission
      Gaia \url{(https://www.cosmos.esa.int/gaia)},  processed  by the Gaia Data  Processing  and  Analysis  Consortium \url{(DPAC, https://www.cosmos.esa.int/web/gaia/dpac/consortium)}. Funding for the DPAC has
      been provided by national institutions, in particular the institutions participating in the Gaia Multilateral Agreement. This research has also made use of the SIMBAD database, operated at CDS, Strasbourg, France. LS \& SG are senior FNRS research associates. 
\end{acknowledgements}

%-------------------------------------------------------------------
\bibliographystyle{aa}
\bibliography{biblio}

\begin{appendix}
\section{ Reliability of the S-star masses}\label{massqualitycheck}
\subsection{Height above the galactic plane}
\begin{figure*}[hbt!]
\begin{centering}
    \mbox{\includegraphics[scale=0.355,trim={0cm 0cm 0cm 0cm}]{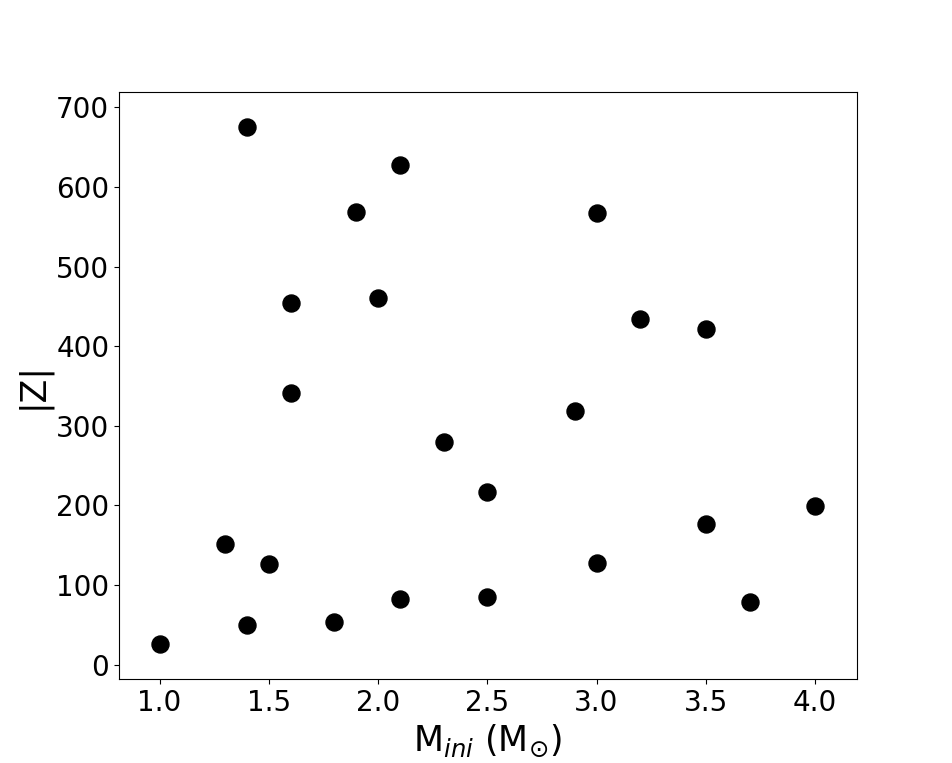}}   
    \hspace{0px}
    \mbox{\includegraphics[scale=0.33,trim={0cm 0cm 0cm 0cm}]{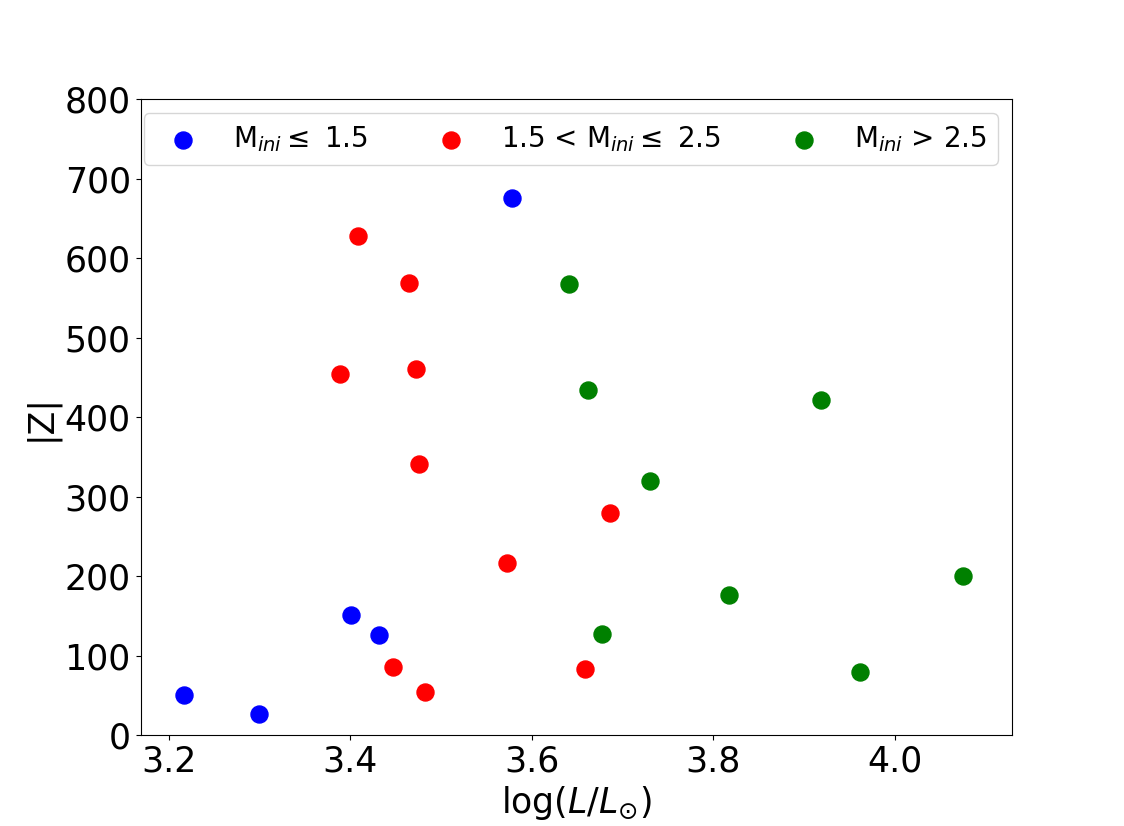}}   
    \caption{\label{Heightgalacticplane}Height above the galactic plane for the sample intrinsic stars and for the intrinsic stars from S18, S19 and S20, as a function of their initial mass (left panel) or luminosity (right panel). }.
\end{centering}
\end{figure*}

Figure~\ref{Heightgalacticplane} presents the height above the galactic plane ($|Z|$) of our sample stars. Though a large $Z$ scatter is present at low masses, the maximum $|Z|$ value decreases with increasing mass, as expected. This trend somehow validates the masses estimated in the present paper.
Nevertheless, the limited size of the present sample, and the absence of bias control in this sample, precludes us from drawing any further conclusion.

\subsection{Gaia-2MASS photometry}
\begin{figure}
    \centering
    \includegraphics[scale=0.3]{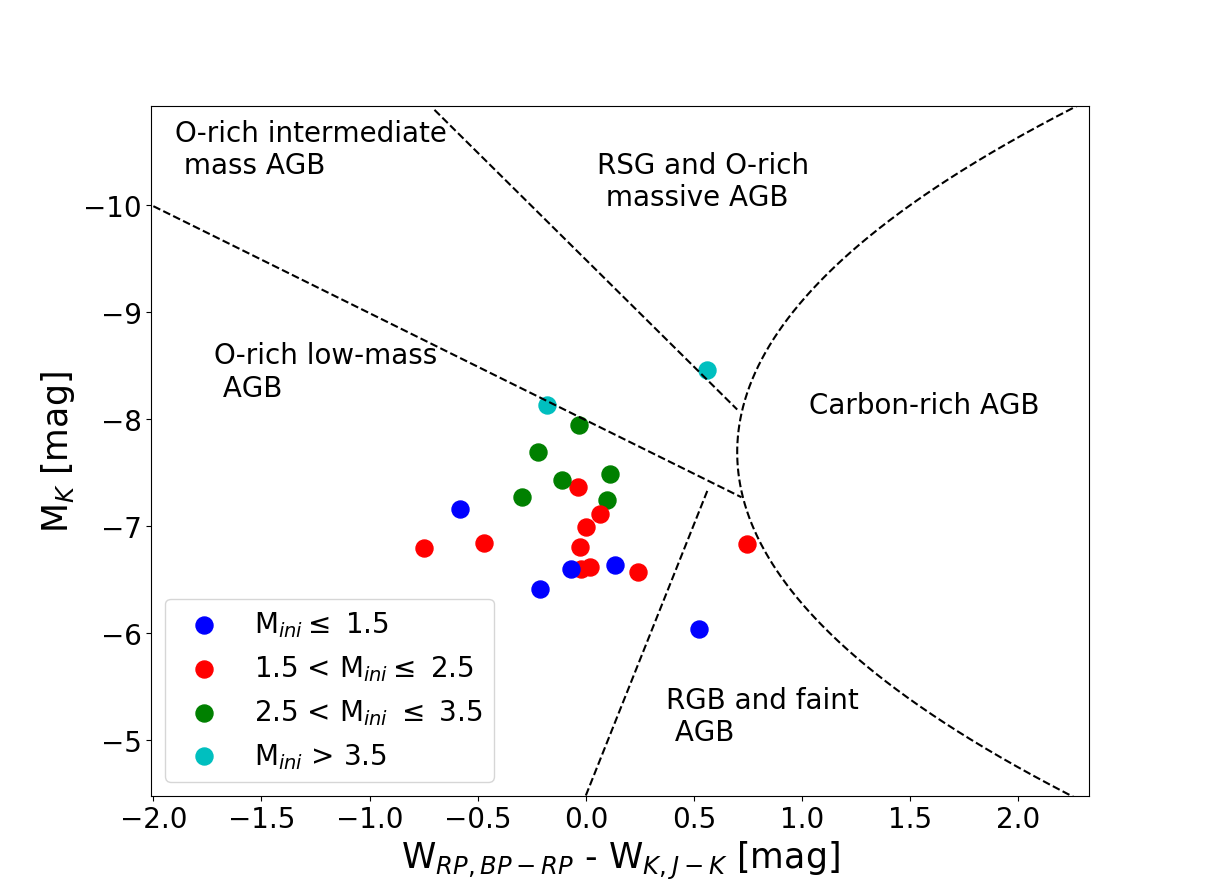}
    \caption{The $W_{RP, BP-RP} - W_{K, J-K}$ vs $M_{K}$ diagram of our sample of Tc-rich S stars.  The definitions of the boundaries (black dashed lines) are from \cite{lebzelter2018}.}
    \label{Gaia2MASS}
\end{figure}

\cite{lebzelter2018} present a classification of AGB stars using Gaia and 2MASS photometry, in the plane $W_{RP, BP-RP} - W_{K, J-K}$ vs $M_{K}$, where $M_{K}$ is the absolute $K_S$ magnitude from 2MASS and Gaia EDR3 parallaxes. The Wesenheit functions $W_{RP, BP-RP}$ and $W_{K, J-K}$ are calculated using the definitions from \cite{lebzelter2018}, i.e., $W_{K, J-K} = K_{S} - 0.686\; (J-K)$ and $W_{RP} = G_{RP} - 1.3 \; (G_{BP} - G_{RP})$, with $J$ and $K_S$ from 2MASS, and $G_{BP}$ and $G_{RP}$ are the magnitudes in the Gaia $BP$ and $RP$ bands. In Fig.~\ref{Gaia2MASS}, we plot our sample stars in this classification scheme.
Most of our sample stars with masses smaller than 2.5 M$_\odot$ lie as expected in the "oxygen-rich low-mass AGB" zone. While an oxygen-rich chemistry is indeed expected for S-type stars (which have C/O smaller than unity), their location in the low-mass zone constitutes a nice confirmation of the masses derived in the present work.
The case of HD~357941, which was flagged as a low-mass star ($M_{\rm ini}\sim 1$~\msun) based on DR2 parallaxes in S19, is noteworthy. We re-evaluated its mass with Gaia EDR3 parallaxes and found it to be $M_{\rm ini} = 3.5$~\msun. This revised mass is now consistent with its location in Fig.~\ref{Gaia2MASS} ($M_K = -7.94, W_{RP, BP-RP} - W_{K, J-K} = -0.03$), where it is located at the border between the low-mass and intermediate-mass O-rich AGB stars.

\section{Error analysis of the Tc and Li abundances}
\label{ErrorTc_Li}

Uncertainties on the Li and Tc abundances of V915~Aql are listed in the bottom panel of Table~\ref{LiandTcabunderr}. In the upper panel, model A designates the adopted model for V915~Aql from S18, whereas models B–G correspond to models differing by one grid step from model A, each parameter varied at a time. The abundances resulting from each of these models is then compared with the abundance from model A and these differences are listed as columns $\Delta_{B - A}$, ..., $\Delta_{G - A}$ in the bottom panels of Table~\ref{LiandTcabunderr}. Model H is the one used to compute the error on the abundances of V915~Aql as described in Section~\ref{abundanceuncertainty} (or Section~4.4 of S18). The contribution from the atmospheric parameter uncertainties on the error on the Li and Tc abundances is given by $\Delta_{H - A}$ in Table~\ref{LiandTcabunderr} as described in Sect.~\ref{abundanceuncertainty}.

\begin{table}[ht]
    \centering
    \caption{Sensitivity of the Li and Tc elemental abundances of V915~Aql upon variations of its atmospheric parameters.}
    \label{LiandTcabunderr}
    \begin{tabular}{c c c r c c c}
\hline
Model    & \teff & $\log g$ & [Fe/H] & C/O & [s/Fe] &$\chi_{t}$ \\
         &  (K)      & (cm~s$^{-2}$)         & (dex)  &   & (dex)  &  (km~s $^{-1}$)  \\ 
\hline 
A & 3400 & 0.0 & -0.5 & 0.75 & 0.00 &  2.0 \\
B & 3300 & 0.0 & -0.5 & 0.75 & 0.00 &  2.0 \\
C & 3500 & 0.0 & -0.5 & 0.75 & 0.00 &  2.0 \\
D & 3400 & 1.0 & -0.5 & 0.75 & 0.00 &  2.0 \\
E & 3400 & 0.0 & 0.0  & 0.75 & 0.00 &  2.0 \\
F & 3400 & 0.0 & -0.5 & 0.75 & 0.00 &  1.5 \\
G & 3400 & 0.0 & -0.5 & 0.50 & 0.00 &  2.0 \\
H & 3500 & 1.0 & 0.0 & 0.50 &1.00 &  2.0 \\
\hline
\end{tabular}

\begin{tabular}{c rrrrrrr}
\hline
Element & $\Delta_{B-A}$  & $\Delta_{C-A}$ & $\Delta_{D-A}$&$\Delta_{E-A}$&$\Delta_{F-A}$&$\Delta_{G-A}$&$\Delta_{H-A}$\\
\hline
[Li/Fe] &-0.06 & -0.5 & 0.8 & 0.0 & 0.6 & - & -0.7 \\
\hline
$\log \epsilon_\mathrm{Tc}$ & -0.1 & -0.1 & - & -0.1 & -0.2 & -0.2 & -0.5 \\
\hline
\end{tabular}
\tablefoot{The atmospheric parameters of V915~Aql are adopted from S18. A dash in the $\Delta$ column indicates that the agreement between the observed and the synthetic spectra was too poor and that the (unique) line usually providing the abundance for the considered element could not be used for that purpose. Column $\chi_{t}$ corresponds to the microturbulence.}
\end{table}

\section{Elemental abundances of sample stars}

In this section we present the tables listing the elemental abundances for the stars of our current study. Table~\ref{tab:Liabund} lists the Tc and Li abundances in our sample stars and also in Tc-rich stars from S18, S19 and S20 which we computed during the current study. Tables~\ref{abundance1}, \ref{abundance2}, and \ref{abundance3} provide the full list of elemental abundances in our sample stars. Table~\ref{hslsandsindex} lists the different elemental indices [ls/Fe], [hs/Fe], [hs/ls], and [s/Fe] for our sample stars as well as for Tc-rich stars from S18, S19, and S20. 

\begin{table}[]
\caption{Tc and Li abundances of the sample stars as well as of the stars from S18, S19 and S20. }
    \label{tab:Liabund}
    \centering
    \begin{tabular}{l|r r r}
    \hline
    Star &$\log \epsilon_\mathrm{Tc}$& $\log \epsilon_\mathrm{Li}$& [Li/Fe] \\
    \hline
         HR Peg & -1.3 & 1.3  & 0.6 \\
         HD 28833 & -1.2 & < -0.3  & < -1.05  \\
         V812 Oph & 1.0 & - & - \\
         AA Cam & -0.4 & < -0.3  & < -1.02\\
         V1139 Tau & -0.6 & -  & - \\
         KR CMa & - & - & -  \\
         CSS~151 & -0.1 & < -0.6 & < -1.4 \\
         CD -27\deg5131 & - & - & -  \\
         BD -18\deg2608 & -0.1 & < 0.4  & < -0.34 \\
         V679 Oph & 0.5 & 0.0 & -0.53 \\
         CSS 454 & 0.5 & < -0.6 & < -1.26 \\
         CSS~1152 & 0.5 & 0.0 & -0.91 \\
         HD 64147 & -0.1 & < -0.6  & < -0.98 \\
    \hline 
    \omiori & -0.4 & < -0.6  & < -1.37 \\
    BD +79\deg156 & -0.7 & < 0.1 & < -0.78 \\
    V915 Aql & -0.2 & < -0.6 & < -1.15 \\
    UY Cen & - & -0.3 & -1.35 \\
    NQ Pup & -0.3 & < -0.6  & < -1.35 \\
    HD 357941 & -0.6 & < -0.2  & < -0.77 \\
    CSS 154 & -0.6 & < -0.3  & < -1.06  \\
    CSS 182 & 0.0 & < -0.6  & < -1.25  \\
    CD -29\deg5912 & -0.6 & < -0.9 & < -1.55 \\
    BD +34\deg1698 & -0.6 & < -0.2 & -0.55 \\
    \hline
    \end{tabular}
 \tablefoot{A dash indicates that the agreement between the observed and the synthetic spectra was too poor and that the (unique) line usually providing the abundance for the considered element had to be rejected. The "<" symbol is used when the line is still usable but only provides an upper limit on the abundance. }   
\end{table}

\begin{table*}
\caption{\label{abundance1} Elemental abundances of sample stars, along with the standard deviation due to line-to-line scatter. }
\centering
\begin{tabular}{lll | lrrrl | lrrrl}
\hline
 & & &\multicolumn{5}{c}{HR Peg}&\multicolumn{5}{c}{HD 288833}\\
\hline
      & Z  & $\log {\epsilon}_\odot$ & $\log \epsilon$ & N   & [X/H] & [X/Fe] & $\sigma_{\rm[X/Fe]}$  & $\log \epsilon$ & N   & [X/H] & [X/Fe] & $\sigma_{\rm[X/Fe]}$  \\
      \hline
C     & 6  & 8.43  & 8.05 &  -   & -0.37    & -0.02     & 0.17           & 8.05 &  -  & -0.37    & -0.06     &   0.17\\
N     & 7  & 7.83  & 8.45 $\pm$ 0.10  &  -   & 0.62      & 0.97       & 0.64        & 8.60 $\pm$ 0.13   & -   & 0.77      & 1.07      & 0.64 \\
O     & 8  & 8.69  & 8.36  & -    & -0.33     & 0.02       & -            & 8.36  &  -   & -0.33     & -0.02     &- \\
Fe    & 26 & 7.50  & 7.15 $\pm$ 0.18  & 14  & -0.35     & -         & 0.22        & 7.19 $\pm$ 0.14 & 13 & -0.30    & -          & 0.19 \\
%Sr I  & 38 & 2.87  & -     & -   & -  & -   & -            & 2.55 $\pm$ 0.21  & 2 & -0.32     & -0.016     & 0.23 \\
Y I   & 39 & 2.21  & 2.30 $\pm$ 0.00  & 1   & 0.09     & 0.44     & 0.25       & 2.50 $\pm$ 0.00 & 1  & 0.29    & 0.59     & 0.19\\
Y II  & 39 & 2.21  & 1.90 $\pm$ 0.00  & 1   & -0.31     & 0.04       & 0.25         & -   & -  & -& - & -\\
Zr I  & 40 & 2.58  & 2.85 $\pm$ 0.21  & 2   & 0.27      & 0.62       & 0.23        & 2.75 $\pm$ 0.07 & 2  & 0.17      & 0.47     & 0.12\\
Nb I  & 41 & 1.46  & 1.35 $\pm$ 0.21  & 2   & -0.11     & 0.24       & 0.23        & 1.30 $\pm$ 0.17  & 3  & -0.16     & 0.14      & 0.19\\
Tc I & 43 &- & -1.30 $\pm$ 0.00  & 1  &  -   & -      & -        & -1.20 $\pm$ 0.00  & 1  &  -   & -     & -  \\
Ba I  & 56 & 2.18  & 1.90 $\pm$ 0.00   & 1   & -0.28     & 0.07       & 0.17        & -     &    & - & -   &- \\
Ce II & 58 & 1.58  & 1.22 $\pm$  0.25  & 5   & -0.36     & -0.01:     & 0.17        & 1.38 $\pm$ 0.29  & 6  & -0.20     & 0.10      & 0.33\\
Pr II & 59 & 0.72  & 1.00 $\pm$ 0.00   & 1   & 0.28      & 0.63       & 0.14        & 1.05 $\pm$ 0.21  & 2  & 0.33      & 0.63     & 0.22\\
Nd II & 60 & 1.42  & 1.53 $\pm$ 0.05  & 3   & 0.11      & 0.46       & 0.22        & 1.42 $\pm$ 0.15 & 4  & 0.01    & 0.31      & 0.26\\
%Sm II & 62 & 0.96  & 0.6 $\pm$ 0.0   & 1   & -0.36     & -0.01      & 0.14        & -     & -   & -  & - & -\\
Eu II & 63 & 0.52  & 0.20 $\pm$ 0.00  & 2 & -0.32     & 0.03       & 0.14        & 0.60 $\pm$ 0.14   & 2  & 0.08      & 0.38      & 0.17 \\   
\hline\\
\end{tabular}
%\end{table*}

%\begin{table*}
%\caption{AA Cam and V1139 Tau}
\centering
\begin{tabular}{lll | lrrrl | lrrrl}
\hline
 & & &\multicolumn{5}{c}{AA Cam}&\multicolumn{5}{c}{V1139 Tau}\\
 \hline
      & Z  & $\log {\epsilon}_\odot$ & $\log \epsilon$ & N   & [X/H] & [X/Fe] & $\sigma_{\rm[X/Fe]}$  & $\log \epsilon$ & N   & [X/H] & [X/Fe] & $\sigma_{\rm[X/Fe]}$  \\
      \hline
C     & 6  & 8.43  & 8.06 & -   & -0.37    & -0.04     & 0.25     & 8.53 & -   & 0.11     & 0.17      &  0.17           \\
N     & 7  & 7.83  & 8.40 $\pm$  0.13 & -   & 0.57      & 0.89      & 0.65        & 9.90 $\pm$ 0.15 & -   & 2.07      & 2.13      & 0.65       \\
O     & 8  & 8.69  & 8.36  & -   & -0.33     & 0.00     &  -           & 8.66  &    & -0.03     & 0.03      &            \\
Fe    & 26 & 7.50  & 7.17 $\pm$ 0.12 & 15 & -0.33    & -         & 0.18        & 7.44 $\pm$ 0.13 & 14 & -0.06    & -          & 0.19       \\
Y I   & 39 & 2.21  & 1.90 $\pm$ 0.00  &1  & -0.31     & 0.02     & 0.22        & -   &-& -     & -  & -     \\
Y II  & 39 & 2.21  & 2.20 $\pm$ 0.00  & 1  & -0.01     & 0.32      & 0.22        & -     &-    & - & -&  -\\
Zr I  & 40 & 2.58  & 2.75 $\pm$ 0.21   & 2  & 0.17      & 0.49      & 0.23        & 2.80 $\pm$ 0.00  & 1  & 0.22      & 0.28      & 0.14       \\
Nb I  & 41 & 1.46  & 1.30 $\pm$ 0.17  & 3  & -0.16     & 0.16      & 0.19        & 1.35 $\pm$ 0.21  & 2  & -0.11     & -0.05     & 0.23       \\
Tc I & 43 & - & -0.40 $\pm$ 0.00  & 1  &    - & -      &-         & -0.60 $\pm$ 0.00  & 1  &-     &-      &-  \\
Ba I  & 56 & 2.18  & 1.90 $\pm$ 0.0  & 1  & -0.28     & 0.05      & 0.17        & 2.40 $\pm$ 0.00  & 1  & 0.22      & 0.28      & 0.17       \\
Ce II & 58 & 1.58  & 1.24 $\pm$ 0.28  & 6  & -0.34    & -0.01:    & 0.15        & 1.90 $\pm$ 0.10   & 3  & 0.32      & 0.38      & 0.17       \\
Nd II & 60 & 1.42  & 1.30 $\pm$ 0.17   & 3  & -0.12     & 0.21      & 0.17        & -     &   - & -& -  & -            \\
\hline\\
\end{tabular}
%\end{table*}

\begin{tabular}{lll | lrrrl | lrrrl}
\hline
 & & &\multicolumn{5}{c}{KR CMa}&\multicolumn{5}{c}{CSS~151}\\
 \hline
      & Z  & $\log {\epsilon}_\odot$ & $\log \epsilon$ & N   & [X/H] & [X/Fe] & $\sigma_{\rm[X/Fe]}$  & $\log \epsilon$ & N   & [X/H] & [X/Fe] & $\sigma_{\rm[X/Fe]}$  \\
      \hline
C     & 6  & 8.43  & 8.36 & -   & -0.07    & -0.06     &    0.17         & 8.06 &  -  & -0.37    & -0.11     &  0.26           \\
N     & 7  & 7.83  & 8.80 $\pm$ 0.18   & -   & 0.97      & 0.98      & 0.66        & 8.60 $\pm$ 0.12  & -   & 0.77      & 1.03      & 0.65        \\
O     & 8  & 8.69  & 8.66  & -   & -0.03     & -0.01     &      -       & 8.36  & -   & -0.33     & -0.07     &             \\
Fe    & 26 & 7.50  & 7.48 $\pm$ 0.18 & 14 & -0.01    &-      & 0.22        & 7.24 $\pm$ 0.14 & 14 & -0.26    & 0          & 0.19        \\
Y I   & 39 & 2.21  & 2.80 $\pm$ 0.00  & 1  & 0.59      & 0.60     & 0.14        & 2.50 $\pm$ 0.00  & 1  & 0.29     & 0.54     & 0.31        \\
Zr I  & 40 & 2.58  & -     & -  & -         & -          &  -           & 2.95 $\pm$ 0.21  & 2  & 0.37      & 0.63      & 0.23        \\
%Nb I  & 41 & 1.46  & - & -  & -      & -      & 0.14        & -     & -   & - & -   &    -         \\
Tc I & 43 &- &-  &-   &-     &-       &-         & -0.10 $\pm$ 0.00  & 1  & -    & -     & - \\
Ba I  & 56 & 2.18  & 2.50 $\pm$ 0.00   & 1  & 0.32      & 0.33      & 0.17        & 2.20 $\pm$ 0.00   & 1 & 0.02      & 0.28      & 0.17        \\
Ce II & 58 & 1.58 & 1.70 $\pm$ 0.30  & 4  & 0.12 & 0.13      & 0.22        & 1.81 $\pm$ 0.28  & 6  & 0.23      & 0.48      & 0.17        \\
Pr II & 59 & 0.72  & -     &-    & -  & -   &-             & 0.90 $\pm$ 0.00   & 1  & 0.18      & 0.43      & 0.14        \\
Nd II & 60 & 1.42  & 2.30 $\pm$ 0.00   & 1  & 0.88      & 0.89      & 0.24        & 1.57 $\pm$ 0.30 & 4  & 0.15     & 0.41      & 0.37        \\
Sm II & 62 & 0.96  & -     & -   & -  & -   & -            & 1.10 $\pm$ 0.00   & 1  & 0.14      & 0.40      & 0.14        \\
Eu II & 63 & 0.52  & -     &  -  & -  & -   & -            & 0.50 $\pm$ 0.00   & 1  & -0.02     & 0.24      & 0.14    \\
\hline\\
\end{tabular}
\tablefoot{Solar abundances (third column) are from \cite{asplund2009}. The  column labelled $N$ lists the number of lines used to derive the corresponding abundance. The $\sigma_{\rm [X/Fe]}$ column lists the total uncertainty on the abundance calculated using the method described in Sect.~\ref{abundanceuncertainty}. A colon flags an uncertain value.}
\end{table*}

\begin{table*}
\caption{\label{abundance2}Same as Table~\ref{abundance1}}
\centering
%\begin{table*}
%\caption{CD -27 5131 and BD -18 2608}
\centering
\begin{tabular}{lll | lrrrl| lrrrl}
\hline
 & & &\multicolumn{5}{c}{CD -27\deg5131}&\multicolumn{5}{c}{BD -18\deg2608}\\
 \hline
      & Z  & $\log {\epsilon}_\odot$ & $\log \epsilon$ & N   & [X/H] & [X/Fe] & $\sigma_{\rm[X/Fe]}$  & $\log \epsilon$ & N   & [X/H] & [X/Fe] & $\sigma_{\rm[X/Fe]}$  \\
      \hline
C     & 6  & 8.43  & 8.06 & -   & -0.37    & -0.07     & 0.26            & 8.23 &-    & -0.19    & 0.11      &   0.10          \\
N     & 7  & 7.83  & 9.60 $\pm$ 0.11   & -   & 1.77      & 2.07       & 0.64        & 8.60 $\pm$ 0.12   &  -  & 0.77      & 1.08       & 0.64        \\
O     & 8  & 8.69  & 8.36  & -   & -0.33     & -0.03      &    -         & 8.36  & -   & -0.33     & -0.02      &   -          \\
Fe    & 26 & 7.50   & 7.20 $\pm$ 0.12   & 10 & -0.30     & - & 0.18        & 7.18 $\pm$ 0.16 & 13 & -0.31     & -          & 0.21        \\
Sr I  & 38 & 2.87  & -     & -   & -  & -   &  -           & 3.00 $\pm$ 0.00     & 1  & 0.13      & 0.44       & 0.14        \\
Y I   & 39 & 2.21  & 2.80 $\pm$ 0.00   & 1  & 0.59      & 0.89       & 0.19        & 2.40 $\pm$ 0.00   & 1  & 0.19     & 0.50          & 0.4         \\
Y II  & 39 & 2.21  & 2.60 $\pm$ 0.00   & 1  & 0.39      & 0.69       & 0.19        & 2.50 $\pm$ 0.00   & 1  & 0.29      & 0.60  & 0.40  \\
Zr I  & 40 & 2.58  & 2.95 $\pm$ 0.21  & 2  & 0.37      & 0.67       & 0.23        & 3.00 $\pm$ 0.14    & 2  & 0.42      & 0.73  & 0.17  \\
Nb I  & 41 & 1.46  & 1.20 $\pm$ 0.00   & 1  & -0.26     & 0.04       & 0.14        & 1.50 $\pm$ 0.00   & 1  & 0.04     & 0.35      & 0.14        \\
Tc I & 43 &- & -  & -  &  -   &   -    &  -       & -0.10 $\pm$ 0.00  & 1  &   -  &   -   & - \\
Ba I  & 56 & 2.18  & 2.80 $\pm$ 0.00   & 1  & 0.62      & 0.92       & 0.17        & 2.50 $\pm$ 0.00   & 1  & 0.32      & 0.63       & 0.17 \\
Ce II & 58 & 1.58  & 1.90 $\pm$ 0.25  & 7  & 0.32       & 0.62        & 0.25        & 2.14 $\pm$ 0.17  & 7  & 0.56      & 0.87      & 0.22        \\
Nd II & 60 & 1.42  & 1.80 $\pm$ 0.00   & 1  & 0.38      & 0.68       & 0.14        & 1.95 $\pm$ 0.21  & 2  & 0.53      & 0.84       & 0.23        \\
Sm II & 62 & 0.96  & 1.00 $\pm$ 0.00     & 1  & 0.04      & 0.34       & 0.14        & 1.30 $\pm$ 0.00   & 2  & 0.34      & 0.65       & 0.14        \\
Eu II & 63 & 0.52  & -     &-    & - &-  &    -         & 0.50 $\pm$ 0.00   & 1  & -0.02     & 0.29       & 0.14    \\
\hline\\
\end{tabular}
%\end{table*}

%\begin{table*}
%\caption{V679 Oph and BD +79 156}
\centering
\begin{tabular}{lll | lrrrl | lrrrl}
\hline
 & & &\multicolumn{5}{c}{V679 Oph}&\multicolumn{5}{c}{V812 Oph}\\
 \hline
      & Z  & $\log {\epsilon}_\odot$ & $\log \epsilon$ & N   & [X/H] & [X/Fe] & $\sigma_{\rm[X/Fe]}$  & $\log \epsilon$ & N   & [X/H] & [X/Fe] & $\sigma_{\rm[X/Fe]}$  \\
      \hline
C     & 6  & 8.43  & 8.31 & -  & -0.11    & 0.40      &  0.10           & 8.06 & -   & -0.37    & 0.00 & 0.25\\
N     & 7  & 7.83  & 8.50 $\pm$ 0.09   & -  & 0.67      & 1.19       & 0.64        &10.10 $\pm$ 0.07 &  -  & 2.27      & 2.64     & 0.65        \\
O     & 8  & 8.69  & 8.36  & -  & -0.33     & 0.19       &       -      &  8.36  & -   & -0.33     & 0.04      &   -          \\
Fe    & 26 & 7.50   & 6.98 $\pm$ 0.14  & 9 & -0.52     & -        & 0.19        & 7.12 $\pm$ 0.13  & 12 & -0.37   & -       & 0.19        \\
Sr I  & 38 & 2.87  & 3.40 $\pm$ 0.00   & 1 & 0.53      & 1.05       & 0.14            & -  & - & -   & -       &      \\
Y I   & 39 & 2.21  & 3.10 $\pm$ 0.00   & 1 & 0.89      & 1.41       & 0.17        &  - & -  & -     & -     & -      \\
Zr I  & 40 & 2.58  & 3.60 $\pm$ 0.34   & 2 & 1.02      & 1.54       & 0.36        & 3.25 $\pm$ 0.21 & 2  & 0.67      & 1.04      & 0.23        \\
Nb I  & 41 & 1.46  & 1.50 $\pm$ 0.00  & 1 & 0.04     & 0.56       & 0.36        & 1.50 $\pm$ 0.00   & 1  & 0.04     & 0.41      & 0.14        \\
Tc I & 43 & & 0.50 $\pm$ 0.00  & 1  &  -   &    -   & -        &  1.00 $\pm$ 0.00  & 1  & -    &    -  &-  \\
Ba I  & 56 & 2.18  & 3.10 $\pm$ 0.00   & 1 & 0.92      & 1.44       & 0.17        & 2.80 $\pm$ 0.00   & 1  & 0.62      & 0.99    & 0.17        \\
Ce II & 58 & 1.58  & 2.35 $\pm$ 0.23  & 7 & 0.77      & 1.29       & 0.26        &  2.17 $\pm$ 0.25  & 7  & 0.59      & 0.96     & 0.15        \\
Pr II & 59 & 0.72  & 2.10 $\pm$ 0.00   & 1 & 1.38      & 1.90        & 0.14        &  -     & -   & - & -   &- \\
Nd II & 60 & 1.42  & 2.40 $\pm$ 0.17   & 3 & 0.98      & 1.50        & 0.19        & 2.30 $\pm$ 0.34   & 3  & 0.88      & 1.25     & 0.4         \\
%Sm II & 62 & 0.96  & -     &-   & -& -& -            & 
%Eu II & 63 & 0.52  & -     & -  & -& -   & -            & 
\hline\\
\end{tabular}

\begin{tabular}{lll | lrrrl | lrrrl}
\hline
 & & &\multicolumn{5}{c}{CSS 454}&\multicolumn{5}{c}{CSS~1152}\\
 \hline
      & Z  & $\log {\epsilon}_\odot$ & $\log \epsilon$ & N   & [X/H] & [X/Fe] & $\sigma_{\rm[X/Fe]}$  & $\log \epsilon$ & N   & [X/H] & [X/Fe] & $\sigma_{\rm[X/Fe]}$  \\
      \hline
C     & 6  & 8.43  & 8.06 &  -  & -0.37    & 0.02      &   0.25          & 8.64 & -  & 0.21     & 0.35      &  0.10           \\
N     & 7  & 7.83  & 9.30 $\pm$ 0.14  &  -  & 1.47      & 1.86       & 0.65        & 8.60 $\pm$ 0.05   & -  & 0.77      & 0.91       & 0.63        \\
O     & 8  & 8.69  & 8.36  & -   & -0.33     & 0.06       &      -      & 8.66  & -  & -0.03     & 0.11       &    -         \\
Fe    & 26 & 7.50   & 7.11 $\pm$ 0.14  & 12 & -0.39     & -         & 0.19        & 7.36 $\pm$ 0.13  & 6 & -0.14     & - & 0.19 \\
Y I   & 39 & 2.21  & 2.50 $\pm$ 0.00   & 1  & 0.29     & 0.68& 0.36  & 3.40 $\pm$ 0.00   & 1 & 1.19      & 1.33 & 0.17        \\
Y II  & 39 & 2.21  & 2.50 $\pm$ 0.00   & 1  & 0.29      & 0.68       & 0.36        & -     & - & -  & -   &-             \\
Zr I  & 40 & 2.58  & 3.15 $\pm$ 0.07  & 2  & 0.57      & 0.96       & 0.12        & 3.75 $\pm$ 0.07  & 2 & 1.17      & 1.31       & 0.12\\
Nb I  & 41 & 1.46  & 1.20 $\pm$ 0.00   & 1  &  -0.26    & -0.26      &  0.13  & 2.10 $\pm$ 0.00   & 1 & 0.64      & 0.78       & 0.14        \\
Tc I & 43 & & 0.50 $\pm$ 0.00  & 1  & -    &    -   & -        & 0.50 $\pm$ 0.00  & 1  &    - & -     & - \\
Ba I  & 56 & 2.18  & 2.80 $\pm$ 0.00   & 1  & 0.62      & 1.01       & 0.17        & 3.10 $\pm$ 0.00   & 1 & 0.92      & 1.06       & 0.17\\
Ce II & 58 & 1.58  & 2.17 $\pm$ 0.21  & 8  & 0.59       & 0.98       & 0.25        & 2.66 $\pm$ 0.22   & 3 & 1.08      & 1.22       & 0.24        \\
Pr II & 59 & 0.72  & 1.00 $\pm$ 0.00     & 1  & 0.28      & 0.67       & 0.14        & -     & - & -  & -   & -            \\
Nd II & 60 & 1.42  & 2.10 $\pm$ 0.00   & 1  & 0.68      & 1.07       & 0.14        & 2.75 $\pm$ 0.07  & 2 & 1.33      & 1.47       & 0.12        \\
Sm II & 62 & 0.96  & 1.50 $\pm$ 0.00   & 1  & 0.54      & 0.93       & 0.14        & -     & - & -  & -   &-             \\
Eu II & 63 & 0.52  & 0.55 $\pm$ 0.07  & 2  & 0.03      & 0.42       & 0.12        & -     & - & -  & -   &-            \\
\hline\\
\end{tabular}
\end{table*}

\begin{table*}
\caption{\label{abundance3}Same as Tables \ref{abundance1} and \ref{abundance2}}
\centering
\begin{tabular}{lll | lrrrl}
\hline
 & & &\multicolumn{5}{c}{HD 64147}\\
 \hline
     & Z  & $\log {\epsilon}_\odot$ & $\log \epsilon$ & N   & [X/H] & [X/Fe] & $\sigma_{\rm[X/Fe]}$  \\
\hline
C     & 6  & 8.43  & 8.06 & -   & -0.37 & 0.30      &  0.17           \\
N     & 7  & 7.83  & 8.00 $\pm$ 0.07 & -   & 0.17      & 0.84      & 0.64        \\
O     & 8  & 8.69  & 8.36  & -   & -0.33     & 0.34     &   -          \\
Fe    & 26 & 7.50   & 6.82 $\pm$ 0.10  & 11 & -0.67    & -          & 0.17        \\
Y I   & 39 & 2.21  & 1.90 $\pm$ 0.00  & 1  & -0.31     & 0.36      & 0.17        \\
Y II  & 39 & 2.21  & 1.90  $\pm$ 0.00 & 1  & -0.31     & 0.36       & 0.17        \\
Zr I  & 40 & 2.58  & 2.45 $\pm$ 0.21 & 2  & -0.13     & 0.54      & 0.23        \\
Nb I  & 41 & 1.46  & 1.05 $\pm$ 0.21 & 2  & -0.41     & 0.26      & 0.23        \\
Tc I & 43 & & -0.10 $\pm$ 0.00  & 1  &  -   & -      &      -   \\
Ba I  & 56 & 2.18  & 1.90 $\pm$ 0.00  & 1  & -0.28     & 0.39      & 0.17        \\
Ce II & 58 & 1.58  & 1.21 $\pm$0.20  & 7  & -0.37     & 0.30     & 0.24        \\
Pr II & 59 & 0.72  & 0.40 $\pm$ 0.00   & 1  & -0.32     & 0.35     & 0.14        \\
Nd II & 60 & 1.42  & 1.30 $\pm$ 0.14  & 2  & -0.12     & 0.55      & 0.17        \\
Sm II & 62 & 0.96  & 0.75 $\pm$ 0.07 & 2  & -0.21     & 0.46      & 0.12        \\
Eu II & 63 & 0.52  & 0.20 $\pm$ 0.00  & 1  & -0.32     & 0.35      & 0.14       \\
\hline
\end{tabular}
\end{table*}

\begin{table*}[]
 \caption{The heavy (hs) and light (ls) s-process indices of the sample stars.  }
    \label{hslsandsindex}
    \centering
    \begin{tabular}{l|cccccc}
    \hline
    Star & [ls/Fe] & [hs/Fe] & [hs/ls] & $\sigma_{\rm[hs/ls]}$ & [s/Fe] & $\sigma_{\rm[s/Fe]}$ \\
    \hline
         HR Peg & 0.43 &0.26 & -0.16 & 0.27 & 0.31 & 0.31 \\
         HD 288833 & 0.53 & - & - & - &  - & -  \\
         V812 Oph  & - & 1.12 & - & - & - & -  \\
         AA Cam  & 0.33 &0.12 & -0.20 & 0.32 & 0.24 & 0.36 \\
         V1139 Tau  & - & - & - & - & - & -\\
         KR CMa  & - & 0.61 & - & - &  -  & -\\
         CSS~151  & 0.58 & 0.34 & -0.24 & 0.38 & 0.48 & 0.42 \\
         CD -27\deg5131  & 0.73 & 0.80 & 0.07 & 0.26 & 0.79 & 0.35 \\
         BD -18\deg2608  & 0.64 & 0.73 & 0.09 & 0.26 & 0.63 & 0.29 \\
         V679 Oph  & 1.47 &1.47  & -0.01 & 0.32 & 1.46 & 0.43 \\
         CSS 454  & 0.82 & 1.04 & 0.22 & 0.21 & 0.88 & 0.27 \\
         CSS 1152  &1.32 & 1.26 & -0.05 & 0.20 & 1.23 & 0.27 \\
         HD 64147  &0.45 & 0.47 & 0.02 &0.26 & 0.43  & 0.33\\
    \hline
 \omiori  &0.16 & -0.02 & -0.19 & 0.30 & 0.01 & 0.36 \\
BD +79\deg156  & 0.80 & 0.72 & -0.08 & 0.21 & 0.73 & 0.27\\
    V915 Aql & 0.28 & 0.47 & 0.19 & 0.27 & 0.36 & 0.37\\
    UY Cen & 1.12 & 1.07 & -0.04 & 0.57 & 0.92 & 0.85\\
    NQ Pup & 0.44 & 0.46 & 0.02 & 0.40 & 0.40 & 0.46\\
    HD 357941 & 0.21 & - & - & - & 0.24& 0.75\\
    CSS 154 & 0.44 & - & - & - & 0.40 & 0.87\\
    CSS 182 & 0.63 & 0.70 & 0.07 & 0.62 & 0.66 & 0.82 \\ 
    CD -29\deg5912 & 0.66 & 0.82 & 0.17 & 0.56 & 0.67 & 0.75\\
    BD+34\deg1698 & 0.80 & 0.9 & 0.09 & 0.40 & 0.87 & 0.59\\
    \hline
    \end{tabular}
  \tablefoot{The [ls/Fe] index has been derived using Y and Zr, whereas the [hs/Fe] index relies on Ba and Nd abundances. The error on the [hs/ls] ratio was computed by quadratically adding the error on the individual abundances of the elements considered to compute the [hs/Fe] and [ls/Fe] ratios. The same method is used to estimate the error on the [s/Fe] ratio, which is based on the abundances of Y, Zr and Ba.} 
\end{table*}

\section{Revised stellar parameters and abundances of stars from S18, S19, S20}
As explained in Sect.~\ref{observationalsample}, the Gaia EDR3 parallaxes became available in the course of our current study. Hence, we revised the stellar parameters of Tc-rich and Tc-poor stars from S18, S19 and S20 using the Gaia EDR3 parallaxes. In Table~\ref{parametersrevised} we list these revised stellar parameters which were derived using the same method as described in Sect.~\ref{parametersdetermination}. In Table~\ref{abundtablelowmass} we present the re-computed elemental abundances of BD+34$^\circ$1698 and HD~357941, which are the only two stars for which the EDR3 parallaxes imposed a revision of $\log g$ (see Table~\ref{parametersrevised}).

\begin{table*}
\centering
\caption{\label{parametersrevised} Atmospheric parameters for S stars from S18, S19 and S20. The $L$ column indicates the luminosity and its error due to the Gaia EDR3 error on the parallax. All the other columns are the same as in Table~\ref{finalparams}.}
\begin{tabular}{l c c c r l c c c c }
\hline 
 Name & $T_{\rm eff}$ & $L$ & $\log g$ & [Fe/H] &$\sigma_{\rm [Fe/H]}$& C/O & [s/Fe]  & M$_{\rm curr}$ & M$_{\rm ini}$ \\
  & (K) & ($L_{\odot}$) & & & & & &  ($M_{\odot}$) & ($M_{\odot}$) \\
\hline 
\multicolumn{10}{c}{Tc-rich S stars from S18}  \\
V915 Aql       & 3400  & 2000  & 0  & -0.50  & 0.15 & 0.75 & 0   & 0.8 & 1.0\\
 & (3400; 3400) & (1900; 2000) & (0; 1) & & & (0.65; 0.75) & (0; 1)  & &\\
NQ Pup         & 3700  & 2800 & 1  & -0.30 & 0.05 & 0.50   & 1   & 2.4 & 2.5\\
 & (3500; 3700) & (2700; 2900) & (1; 2) & & & (0.5; 0.75) & (1; 1)  & &\\
UY Cen         & 3300  & 11884 & 0  & -0.30   & 0.15 & 0.999 & 1    &3.6 & 4.0\\
 & (3000; 3400) & (11100; 13000) & (0; 3) & & &(0.971; 0.999) & (1; 2)  & &\\
\hline
\multicolumn{10}{c}{Tc-rich S stars from S19} \\
HD 357941 & 3400 & 8300 & 0 & -0.48 & 0.14 & 0.5 & 0  & 3.2 & 3.5\\
&(100; 100) &(7400; 9300) &(0; 1) & & &(0.5; 0.75) & (0; 1)& &\\
      CSS 154 & 3400 & 3000 & 1 & -0.29 & 0.20  & 0.5 & 0  &1.4 & 1.6\\
       &(100; 100) & (2700; 3300) &(1; 3) & & &(0.5; 0.899) & (0; 1) & & \\
      CSS 182 & 3500 & 2500 & 1 & -0.40 & 0.21  & 0.5 & 1 & 1.1 & 1.3 \\
       &(100; 100)&(2200; 2800) &(1; 3) & &  &(0.5; 0.899) & (1; 1)& &\\
      CD $-29^\circ$5912 & 3600 & 2700 & 1 & -0.40 & 0.22  & 0.5 & 1  & 1.3 & 1.5\\
       &(100; 100) & (2500; 3000)& (1; 3) & &  &(0.5; 0.899) & (1; 1)  & &\\
    BD +34$^\circ$1698 & 3400 & 3800 & 0 & -0.70 & 0.20 & 0.5 & 1  & 1.2 & 1.4 \\
      &(200; 200)& (3300; 4400) & (0; 3) & &  &(0.5; 0.899)&(1; 1)  & &\\
\hline
\multicolumn{10}{c}{Tc-rich S stars from S20}\\
\omiori & 3500 & 3000 & 1  & -0.28  &  0.19  & 0.75 & 0   & 1.6 & 1.8\\
& (3500; 3600) & (2800; 3300)   & (0; 2) & &  & (0.5; 0.90) & (0; 1) & &\\
BD +79 156 & 3600 & 2600 & 1 & -0.16  & 0.12  & 0.50 &  1 & 2 & 2.1\\
& (3600; 3700) & (2400; 2700)  &  (1; 3) & &  & (0.50; 0.75) & (1; 1) & &\\
\hline
\multicolumn{10}{c}{Tc-poor S stars from S18} \\
HD 189581      & 3500  & 1700  & 1  & 0.00   & 0.13 & 0.50   & 0  3 & 1.6&1.8\\
 & (3500;3500) & (1600; 1800) & (1; 2) & & & (0.50; 0.75) &(0; 0) & &\\
HD 233158      & 3600  & 1900  & 1  & -0.40  & 0.16 & 0.50  & 1  & 1.1&1.3\\
 & (3600; 3600) & (1800; 2000) & (1; 1) & & & (0.50; 0.75) & (1; 1) & &\\
HD 191589      & 3700 & 500  & 1 & -0.30  & 0.10 & 0.75  & 1   &1 & 1\\
 & (3700; 3800) &(500; 600)&(1; 2) & & &(0.75;0.75)  &(1; 1)  & &\\
HD 191226      & 3600  & 6400 & 1  & -0.10  & 0.13 & 0.75   & 1   & 4.6 & 5.0\\
 & (3600; 3600) &(6100; 6800) & (0; 1) & & & (0.50; 0.899)&(1; 1) & & \\
V530 Lyr       & 3500 & 1700  & 1  & 0.00  & 0.10 & 0.50   & 1   &1.6&1.8\\
 & (3500; 3600)& (1600; 1800)&(1; 3)& & &(0.50; 0.75)&(1; 1) & & \\
HD 215336      & 3700  & 600  & 1  & 0.00  & 0.12 & 0.50 & 1    &1.0 & 1.2 \\
& (3700; 3700)&(500; 600)&(1; 1)& & & (0.50; 0.75)&(1; 1)& & \\
HD 150922      & 3600  & 5000  & 0  & -0.50 & 0.12 & 0.50 & 1   & 2.1&2.2 \\
 & (3600; 3600) & (4800; 5100) & (0; 1) & & & (0.50; 0.75) & (1; 1) & &\\
HD 63733       & 3700 & 1500  & 1  & -0.10 & 0.13 & 0.50 & 1  &2.2&2.3 \\
 & (3700; 3700) & (1400; 1600) & (1; 2) & & & (0.50; 0.75) & (1; 1) &  &\\
BD +69$^\circ$524     & 3600  & 900  & 1  & -0.40 & 0.11 & 0.50 & 0    &1.0& 1.0\\
 & (3600; 3600) & (800; 900) &(1; 2) & &  &(0.50; 0.75)& (0; 0) & & \\
BD +28$^\circ$4592    & 3700  & 600   & 1  & -0.10 & 0.12 & 0.75 & 1  &1.1& 1.3\\
 & (3700; 3800) & (600; 700) & (1; 2) & & & (0.50; 0.899)&(1; 1) & & \\
V1135 Tau      & 3400  & 1700 & 1  & -0.20 & 0.14 & 0.50 & 1   &1.0& 1.2 \\
 & (3400; 3500) & (1600; 1800) & (0; 1) & & & (0.5; 0.951) & (1; 1)  & & \\
AB Col         & 3500  & 2000  & 1 & 0.00 & 0.14 & 0.50  & 1  & 1.9&2.0\\
 & (3300; 3500) & (1900; 2000) & (1; 2) & & & (0.50; 0.75) & (1; 1) & &\\
TYC 5971-534-1 & 3600  & 800   & 1  & -0.10  & 0.18 & 0.899 & 1  & 1.1&1.3\\
 & (3600; 3600) & (700; 900) & (0; 2) & & &(0.50; 0.899) & (1; 2) & & \\
BD -10$^\circ$1977    & 3500  & 6100  & 0  & -0.50  & 0.15 & 0.50 & 1    &1.5&1.7 \\
 & (3500; 3600) & (5700; 6500) & (0; 1) & & & (0.50; 0.899) & (1; 1)& &\\
FX CMa         & 3500  & 6500  & 1  & 0.00  & 0.14 & 0.971& 1  &4.2&4.5  \\
 & (3500; 3500) & (6100; 7100) & (1; 1) & & & (0.951; 0.971) &(1; 1) & & \\ 
BD-22$^\circ$1742     & 4000  & 600  & 1  & -0.30 & 0.09 & 0.75 & 0  & 1.5&1.7 \\
 & (4000; 4000) & (500; 600) & (1; 5) & & & (0.50; 0.899)&(0; 1)  & & \\
 \hline
\end{tabular}
\end{table*}

\begin{table*}
\caption{\label{abundtablelowmass}Elemental abundances for the sample stars, along with the standard deviation due to line-to-line scatter. Solar abundances (third column) are from \cite{asplund2009}. The  column labelled $N$ lists the number of lines used to derive the abundance. The $\sigma_{\rm [X/Fe]}$ column lists the total uncertainty on the abundances calculated using the method described in Sect.~\ref{abundanceuncertainty}. The abundances and total error budget for V915~Aql were retrieved from S19.}
\centering
\begin{tabular}{lcc|lrrrc|lrrrc}
\hline
 & & &\multicolumn{5}{c}{BD +34$^\circ$1698}&\multicolumn{5}{c}{HD 357941}\\
\hline
      & $Z$  & $\log {\epsilon^a}_\odot$ & $\log \epsilon$  & N       & {[}X/H{]} & {[}X/Fe{]}& $\sigma_{\rm [X/Fe]}$ & $\log \epsilon$  & N        & {[}X/H{]} & {[}X/Fe{]}& $\sigma_{\rm[X/Fe]}$\\
\hline
C     & 6  & 8.43  & 8.059      &  -&  -0.371 & 0.329 &- & 8.059      &-&   -0.371        & 0.109 &- \\
N     & 7  & 7.83  & 9.0      &   -      &  1.17  & 1.87   & -       & 8.0      &  -        &    0.17  &  0.65 &-  \\
O     & 8  & 8.69  & 8.36 &    - &  -0.33       &   0.37  & -      & 8.36      & -         & -0.33          & 0.15 &-  \\
Fe    & 26 & 7.5   & 6.80 $\pm0.27$  & 13      & -0.70     &-& 0.3 & 7.02 $\pm0.14$  & 12       & -0.48     & - &0.2 \\
Y I   & 39 & 2.21  & 2.50 $\pm$ 0.1 & 1 &-0.11 & 0.59 & 0.2 & 2.10 $\pm$ 0.00    & 1        & -0.11     & 0.37  & 0.5  \\
Y II  & 39 & 2.21  &2.20 $\pm$ 0.1 & 1 & -0.01& 0.69 & 0.2 & 2.00 $\pm$ 0.00   & 1        & -0.21     & 0.27&  0.5 \\
Zr I  & 40 & 2.58  & 2.65$\pm$ 0.07   & 2       & 0.07      & 0.77 & 0.4 & 2.20 $\pm$ 0.21 & 2        & -0.38     & 0.10 & 0.4  \\
Nb I  & 41 & 1.46  & 1.50$\pm$ 0.00    & 1       & 0.04     & 0.74 &  0.3  & 1.2$\pm$  0.00    & 2        & -0.26     & 0.22 & 0.3 \\
Ba I  & 56 & 2.18  & 2.5$\pm$ 0.00    & 1       & 0.29      & 0.99 &0.4 & 2.0$\pm$ 0.00   & 1        & -0.18      & 0.30  &0.4 \\
Ce II & 58 & 1.58   & 1.24$\pm$ 0.13  & 5       & -0.34     &   0.36  &  0.3    & 1.07  $\pm$ 0.30  & 4        & -0.51      & -0.03 & 0.3 \\
%Pr II & 59 & 0.72  &-& -& -&-& -&-&-& -&-& -& 0.7 $\pm$ 0.00   & 1 & -0.02     & 0.27 &  0.1  \\
%Nd II & 60 & 1.42 & & & & & 1.9 $\pm$ 0.00 & 1 & 0.48 & 0.75 & 1.9 $\pm$ 0.00 & 1 & 0.48 & 0.77 \\
%Sm II & 62 & 0.96 & & & & & 1.0 $\pm$ 0.00 & 1 & 0.04 & 0.31 & 1.0 $\pm$ 0.00 & 1 & 0.04& 0.33 \\
\hline\\
\end{tabular}
\end{table*}

\section{Spectral windows used for spectral fitting in Sect.~\ref{parametersdetermination}}
Table~\ref{Tab:bands} lists the different spectral windows used in the spectral fitting routine to derive an initial estimate of the stellar parameters.

\begin{table}
\caption[]{\label{Tab:bands}
The spectral windows used to compare observed spectra with MARCS synthetic spectra of S stars. The band and normalisation limits are expressed in nm.
}
\centering
\begin{tabular}{rrrr}
\hline\\
$\lambda_{\rm min}$ & $\lambda_{\rm max}$ & $\lambda_{\rm min}$ & $\lambda_{\rm max}$\\
\cline{1-2} \cline{3-4}
\multicolumn{2}{c}{Band} & \multicolumn{2}{c}{Normalisation}\\
\hline\\
          400.0&420.0&408.0&412.0\\ %& 10001\\                       
          420.0&440.0&432.0&434.0\\ %& 10001\\
          440.0&458.0&456.0&458.0\\ %& 9001 &  TiO bandhead starts at $\lambda$458.4\\   
          458.0&480.0&472.0&476.0\\ %& 11001& TiO bandhead starts at $\lambda$480.4\\   
          480.0&490.0&489.4&489.9\\ %& 5001 \\ 
          490.0&495.0&494.0&495.0\\ %&  2501 & TiO band $\lambda$495.4\\   
          495.0&516.0&515.5&516.0\\ %&  10501 & TiO band $\lambda$516.7\\   
          516.0&530.0&516.0&516.3\\ %&  7001 \\      
          530.0&544.0&543.0&544.5\\ %&  7001 & TiO band $\lambda$544.8   \\   
          544.0&565.0&560.5&562.0\\ %& 10501 \\
          565.0&575.5&573.0&573.7\\% & 5251\\
          575.5&588.0&580.0&582.0\\ %& 6251\\
          588.0&615.0&613.0&615.0\\ %& 13501\\
          615.0&647.0&645.0&647.0\\ %& 16001\\
          647.0&671.5&670.0&671.5\\ %& 12251\\
          671.5&685.0&683.0&685.0\\ %& 6751 \\
          703.0&720.0&703.0&705.0\\ %& 8501 \\
\hline
\end{tabular}
\end{table}

\section{Atomic line list}

Table~\ref{linelist} lists the lines used for the abundance analysis.
\begin{table*}
\centering
%\captionsetup{justification=centering}
\caption{Atomic lines used in this study.}
\label{linelist}
\begin{tabular}{l r r r l}

\hline
Species& $\lambda$ [\AA] & $\chi$ [eV] & $\log gf$ & Reference \\
\hline
Fe I & 7386.334 & 4.913  & -0.268  &    \cite{K07}         \\
& 7389.398 & 4.301 & -0.460 & \cite{K07} \\
 & 7418.667 & 4.143 & -1.376 & \cite{BWL} \\
 & 7421.554   & 4.638     & -1.800       &   \cite{MFW}         \\
 & 7443.022 & 4.186 & -1.820 & \cite{MFW} \\
 & 7461.263 & 5.507 & -3.059 & \cite{K07} \\
 & 7498.530 & 4.143 & -2.250 & \cite{MFW} \\
 & 7540.430 & 2.727 & -3.850 & \cite{MFW} \\
 & 7568.899 & 4.283 & -0.773 & \cite{K07} \\
 & 7583.787 & 3.018 &  -1.885 &    \cite{BWL}  \\
 & 7586.018 & 4.313 & -0.458 & \cite{K07} \\
 & 7832.196 & 4.435 & 0.111 & \cite{K07} \\
 & 7937.139 & 4.313 & 0.225 & \cite{K07} \\
 & 8108.320 & 2.728 & -3.898 & \cite{K07} \\
 & 8239.127 & 2.424       &-3.180        &    \cite{BWL}           \\
 & 8248.129 & 4.371       & -0.887        &   \cite{K07}                 \\ %CSS 997
 & 8621.601 &  2.949 & -2.320   &   \cite{K07}         \\
 & 8616.280 & 4.913      & -0.655  &    \cite{NS}         \\
 & 8698.706 & 2.990 & -3.452 & \cite{K07} \\
 & 8699.454 & 4.955 & -0.380 & \cite{NS} \\
 & 8710.404 & 5.742 & -5.156 & \cite{K07} \\
 & 8729.144 & 3.415 & -2.871 & \cite{K07} \\
 & 8747.425 & 3.018 & -3.176 & \cite{K07} \\
 & 8763.966 & 4.652 & -0.146 & \cite{NS} \\
%Fe II & 7454.035 & 10.562 & -4.130 & \cite{RU} \\
%Sr I & 4962.259 & 1.847 & 0.200 & \cite{GC}\\
Sr I & 4872.488 & 1.798 & -0.060 & \cite{VALD1999}\\ %U \\
   & 7070.070 & 1.847 & -0.030 & \cite{GC}\\
Y I & 8800.588 & 0.000 & -2.240 & \cite{CB} \\
Y II & 7881.881 & 1.839 & -0.570 & \cite{Nil}\\
%Zr I & 7562.129 & 0.623 & -2.170 & \cite{CB} & U\\
 Zr I & 7819.374 & 1.822 & -0.380 & \cite{Zrlines} \\
 & 7849.365 & 0.687 & -1.300 & \cite{Zrlines}\\
Nb I & 5189.186 & 0.130 & -1.394 & \cite{DLa} \\
 & 5271.524 & 0.142 & -1.240 & \cite{DLa} \\
 & 5350.722 & 0.267 & -0.862 & \cite{DLa} \\
Tc I & 4238.190 & 0.000  & -0.550 & \cite{palmeri2005}\\
 & 4262.270 & 0.000 &  -0.350 & \cite{palmeri2005}\\
 & 4297.060 & 0.000 &    -0.190 & \cite{palmeri2005} \\
Ba I & 7488.077 & 1.190 & -0.230 & \cite{MW} \\
%Ce I & 8396.397 & 0.719 & -1.177 &\cite{MC} & U\\ 
Ce II & 7580.913 & 0.327 & -2.120 & \cite{PQWB}\\
& 8025.571 & 0.000 & -1.420  & \cite{MC}  \\
& 8394.514 & 0.265   &  -2.590  &\cite{PQWB}  \\%DRM \\
& 8404.133 & 0.704 & -1.670 & \cite{PQWB} \\ 
 & 8405.254 & 0.295 & -2.100 & \cite{MC} \\
 & 8716.659 & 0.122 & -1.980 & \cite{MC} \\
 & 8769.913 & 0.553 & -2.370 & \cite{PQWB}\\
 & 8772.135 & 0.357 & -1.260 & \cite{PQWB} \\
% & 8777.432 & 0.521 & -2.910 & & U \\
Nd II & 4715.586 & 0.205 & -0.900 & \cite{HLSC} \\
 & 5276.869 & 0.859 & -0.440 & \cite{MC} \\
 & 5293.160 & 0.823 & 0.100 & \cite{HLSC}\\
 & 5319.810 & 0.550 & -0.140 & \cite{HLSC} \\
 & 5385.888 & 0.742 & -0.860 & \cite{MC}  \\
 & 7513.736 & 0.933    &    -1.241 & \cite{MC}  \\
Sm II & 7042.206 & 1.076  & -0.760  & \cite{Xu2003} \\
 & 7051.55 & 0.933  & -0.960   & \cite{MC}  \\
Eu II  & 6437.640 & 1.320 & -1.998 & \cite{MC}\\%TM \\
 & 6645.067 & 1.380 & -1.823 & \cite{MC}\\%TM \\
Pr II & 5219.045 & 0.795 & -0.053 & \cite{MC}\\% U\\
& 5322.772 & 0.483 & -0.141 & \cite{MC} \\%U\\
 
\hline

\end{tabular}

\end{table*}

\end{appendix}

\end{document}